\documentclass[twocolumn]{aastex63}
\usepackage{natbib}
\usepackage{color}
\usepackage{mathtools}
\usepackage{hyperref} %----set up hyper link to citations, equations...
\def    \apjl  		{\rm {ApJL}}

\def    \apjl  		{\rm {ApJL}}

\def	\cm		{\,{\rm {cm}}}
\def	\K		{\,{\rm K}}
\def	\g		{\,{\rm {g}}}
\def	\mum	{\,{\mu \rm{m}}}

\def \bea {\begin{eqnarray}}
\def \ena {\end{eqnarray}}

\def	\B	{{\rm B}}

\def	\C	{{\rm C}}

\def	\cm	{\,{\rm cm}}
\def	\d	{{\rm d}}

\def	\D	{{\rm D}}

\def	\erg	{\,{\rm erg}}

\def	\g	{\,{\rm g}}
\def	\gas	{\,{\rm gas}}
\def	\G	{{\rm G}}

\def	\H	{{\rm H}}

\def	\IR	{{\rm IR}}

\def	\N	{{\rm N}}

\def	\s	{\,{\rm s}}

\def	\AU	{\,{\rm au}}

\def	\V	{{\rm V}}

\def	\rad	{{\rm rad}}
\def	\yr	{{\rm yr}}
\def	\km	{\, {\rm km}}
\def    \gas     	{{\rm gas}}

\begin{document}
\shorttitle{Effect of rotational disruption on radiation pressure}
\shortauthors{Thiem Hoang}

%\title{Revisiting radiation pressure feedback in massive star formation: effects of dust rotational disruption}
\title{Effect of dust rotational disruption by radiative torques on radiation pressure feedback from massive protostars}

\author{Thiem Hoang}
\affiliation{Korea Astronomy and Space Science Institute, Daejeon 34055, Republic of Korea, \href{mailto:thiemhoang@kasi.re.kr}{thiemhoang@kasi.re.kr}}
\affiliation{Korea University of Science and Technology, 217 Gajeong-ro, Yuseong-gu, Daejeon, 34113, Republic of Korea}

\begin{abstract}
Radiation pressure on dust is thought to play a crucial role in the formation process of massive stars by acting against gravitational collapse onto the central protostar. However, dust properties in dense regions irradiated by the intense radiation of massive protostars are poorly constrained. Previous studies usually assume the standard interstellar dust model to constrain the maximum mass of massive stars formed by accretion, which appears to contradict with dust evolution theory. In this paper, using the fact that stellar radiation induces on dust simultaneous radiation pressure and radiative torques, we study the effects of grain rotational disruption by radiative torques (RATs) on radiation pressure and explore its implications for massive star formation. { For this paper, we focus on the protostellar envelope and adopt a spherical geometry.} We find that original large grains of micron-sizes presumably formed in very dense regions can be rapidly disrupted into small grains by RATs due to infrared radiation from the hot dust shell near the sublimation front induced by direct stellar radiation. Owing to the modification in the size distribution by rotational disruption, the radiation pressure opacity can be decreased by a factor of $\sim 3$ from the value expected from the original dust model. However, to form massive stars via spherical accretion, the dust-to-gas mass ratio needs to be reduced by a factor of $\sim 5$ as previously found.
\end{abstract}

\section{Introduction}\label{sec:intro}
Radiation pressure on dust plays a central role in numerous astrophysical processes, including formation and feedback of massive stars (\citealt{1971A&A....13..190L}; \citealt{2007ARA&A..45..565M}; \citealt{Tan:2014em}) and supermassive black holes and active galactic nuclei (AGN) (\citealt{2012ARA&A..50..455F}), and stellar and galactic winds (\citealt{2010ApJ...709..191M}). The radiation pressure depends on dust properties (composition and size distribution), which are poorly known in intense radiation fields.

Massive stars are ubiquitous in the universe, yet its formation mechanism is still hotly debated (see reviews by \citealt{2003astro.ph..6595L}; \citealt{2007ARA&A..45..565M}; \citealt{Zinnecker.2007}; \citealt{Motte.2016bb}). From the early studies on massive star formation, radiation pressure on dust has been suggested to be a major barrier for the formation of massive stars (\citealt{1971A&A....13..190L}; \citealt{1974A&A....37..149K}). Dust in the central hottest region is sublimated due to high temperatures, producing a cocoon or a torus surrounding the central core. Most UV photons of massive protostars are absorbed by dust in a thin layer just beyond the sublimation front and are re-emitted in infrared (IR) radiation. The latter radiation of long wavelengths can penetrate deeper into the protostellar envelope and induce radiation pressure on dust. \cite{1971A&A....13..190L} found radiation pressure force induced by IR dust radiation could exceed gravitational force and halt the accretion when the stellar mass exceeds $M_{\star}\sim 20M_{\odot}$.

\cite{1987ApJ...319..850W} (hereafter WC87) studied in detail the maximum stellar mass implied by radiation pressure. The authors calculated the radiation pressure on dust in the outer envelope induced by photons emitted from a thin shell of hot dust by following dust evolution. The authors found that, to produce massive stars of $M>20M_{\odot}$, the dust-to-gas mass ratio must be reduced by a factor of 4, and the grain size distribution must be modified such that large grains of $a\sim 0.05-0.25\mum$ are depleted in the protostellar envelope.

\cite{2002ApJ...569..846Y} performed two-dimensional radiation-hydrodynamic simulations of the collapse of massive cores, including radiation pressure and non-spherical collapse due to rotation. The authors found that the effects of non-spherical collapse could help to form protostars of 30-40$M_{\odot}$ assuming the standard interstellar dust opacity. However, they were unable to make objects much larger than about 40$M_{\odot}$, even when starting with an initial core of 120$M_{\odot}$; eventually the radiative acceleration halted collapse. Three-dimensional radiation-hydrodynamic simulations by \cite{2009Sci...323..754K} reveal the inefficiency of radiation pressure in halting accretion because of radiation Rayleigh-Taylor instability that allows radiation to escape through low density region ({ see recent reviews by \citealt{Krumholz:2014ia} and \citealt{Tan:2014em}). Hydrodynamic simulations by \cite{Klassen.2016} came to the same conclusion that radiation pressure does not halt the accretion, although the standard grain size distribution of the diffuse ISM is assumed.}

{ Nevertheless, to understand the exact role of radiation pressure feedback in massive star formation, it is necessary to accurately understand dust properties in the envelope (cocoon) surrounding the massive (proto-)star.} Previous studies usually assume the standard interstellar dust model from \cite{1977ApJ...217..425M} (hereafter MRN model) with the maximum grain size of $a_{\rm max}=0.25\mum$ (\citealt{1986ApJ...310..207W}; \citealt{1987ApJ...319..850W}). The assumption of the MRN size distribution is difficult to reconcile with the existence of large grains of micron sizes in dense cores implied by theoretical calculations (e.g., \citealt{2013MNRAS.434L..70H}) and inferred from observations (e.g., \citealt{2010Sci...329.1622P}; \citealt{Lefevre:2020fw}; \citealt{2013ApJ...763...55R}; \citealt{Ysard:2013fn}). Grain growth is also observed toward protostellar disks, a later phase of star formation \citep{2009ApJ...696..841K}. Such large grains are most likely to be porous or have composite structures of ice mantles \citep{Guillet.2020}. Therefore, dust in the protostellar envelope plausibly has similar properties as in prestellar cores or even grows to larger sizes, instead of having the standard MRN distribution. { Dust size distribution is especially important for understanding the radiation feedback in massive star clusters (see a review by \citealt{Krumholz:2019dr}).}

In addition to radiation pressure, stellar radiation is known to induce radiative torques (RATs) on dust grains of irregular shapes (\citealt{1976Ap&SS..43..291D}; \citealt{1996ApJ...470..551D}; \citealt{2007MNRAS.378..910L}). Such RATs act to spin up the grain to suprathermal rotation (\citealt{1996ApJ...470..551D}; \citealt{2009ApJ...695.1457H}). \cite{Hoang:2019da} { realized} that centrifugal stress resulting from such suprathermal rotation can exceed the maximum tensile strength of grain material, resulting in the disruption of the grain into fragments. This new physical mechanism was termed Radiative Torque Disruption (RATD). \cite{2016ApJ...818..133S} first noticed that fluffy grains in the solar system could be disrupted by spin-up due to RATs. Since rotational disruption acts to break loose bonds between the grain constituents, unlike breaking strong chemical bonds between atoms in thermal sublimation, RATD can work with the average interstellar radiation field (see \citealt{2020Galax...8...52H} for a review). The RATD mechanism introduces a new environment parameter for dust evolution, namely local radiation intensity, and is found to be the most efficient mechanism that constrains the upper limit of the size distribution (\citealt{2019ApJ...876...13H}; \citealt{Hoang.2021}). We will show in Section \ref{sec:RATs} that grain size modification by RATD occurs faster than grain acceleration by radiation pressure. Thus, the grain size distribution constrained by RATD should be used for calculations of radiation pressure rather than the original dust size distribution expected for the dense prestellar core or the MRN size distribution.

This paper aims to revisit the radiation pressure problem by considering the effect of RATD on the radiation pressure opacity and study its implications for massive star formation. { To explore the effect of RATD and its resulting radiation pressure opacity, we assume a spherical collapse for the protostellar envelope. Indeed, both observations (see \citealt{Cesaroni.2006}) and numerical simulations (e.g., \citealt{2013ApJ...767L..11K}) establish that massive star formation proceeds with the formation of an accretion disk. However, the exact radius of accretion disks is uncertain, depending on the various effects of turbulence, gravity, magnetic fields, feedback, and non-ideal Magneto-hydrodynamics (MHD) effects. Thus, the detailed modeling is devoted in a followup paper.}

The structure of the present paper is as follows. In Section \ref{sec:pressure} we review radiation pressure and study its dependence on the incident radiation field and grain size distribution. In Section \ref{sec:RATs}, we review radiative torques on dust and the rotational disruption mechanism. In Section \ref{sec:protostar}, we describe the working model of massive protostellar cloud and radiation fields. In Section \ref{sec:results}, we present numerical results for the disruption size by RATD and calculate resulting radiation pressure opacity. In Sections \ref{sec:disc} and \ref{sec:summary}, we discuss our main findings and present conclusions.

\section{Radiation Pressure on Dust}\label{sec:pressure}
Photons carry energy, momentum, and angular momentum (spin). Dust grains exposed to a radiation field experience radiation pressure due to light absorption and scattering, which is known to be important in astrophysics (e.g., \citealt{Spitzer:1949bv}). Moreover, interstellar dust grains are { likely} to have irregular shapes as inferred from interstellar polarization (\citealt{Hall:1949p5890}; \citealt{Hiltner:1949p5856}). When subject to an anisotropic radiation field, such grains experience radiative torques (\citealt{1976Ap&SS..43..291D}; \citealt{1996ApJ...470..551D}; \citealt{2004ApJ...614..781A}). As a result, grains simultaneously experience radiation pressure and radiative torques when irradiated by a radiation beam, which affects grain translational and rotational dynamics. In this section, we first describe radiation pressure and its dependence on grain properties. Radiative torques will be described in Section \ref{sec:RATs}.

\subsection{Radiation pressure cross-section}
Let $u_{\lambda}$ be the spectral energy density of radiation field at wavelength $\lambda$. The energy density of the radiation field is then $u_{\rad}=\int_{0}^{\infty} u_{\lambda}d\lambda$. To describe the strength of a radiation field, let define $U=u_{\rm rad}/u_{\rm ISRF}$ with 
$u_{\rm ISRF}=8.64\times 10^{-13}\erg\cm^{-3}$ being the energy density of the average interstellar radiation field (ISRF) in the solar neighborhood as given by \cite{1983A&A...128..212M}. For a black body of temperature $T$, the radiation spectrum is $u_{\lambda}= B_{\lambda}(T)/c$.

The radiation pressure cross-section efficiency for a spherical grain of size $a$ is defined by
\bea
Q_{\rm pr}(a,\lambda)=Q_{\rm abs}+Q_{\rm sca}(1-\langle\cos\theta\rangle),
\ena
where $\langle\cos\theta\rangle$ is the mean cosine of scattering angle $\theta$, and $Q_{\rm abs}=C_{\rm abs}/\pi a^{2}$, $Q_{\rm sca}=C_{\rm sca}/(\pi a^{2})$ are the absorption and scattering efficiency.

The radiation force on a grain is calculated as
\bea
F_{\rm rad,a}=\int_{0}^{\infty}  Q_{\rm pr}(a,\lambda)\pi a^{2}=u_{\rm rad}\bar{Q}_{\rm pr}\pi a^{2}u_{\lambda} d\lambda,
\ena
where the average cross-section efficiency is given by
\bea
\bar{Q}_{\rm pr}(a)=\frac{\int_{0}^{\infty} Q_{\rm pr}(a,\lambda)u_{\lambda}d\lambda}{u_{\rm rad}}.\label{eq:Qpr_avg}
\ena
 
%Above, we assume that photons of $E>E=13.6ev$) are rapidly absorbed by H atoms and choose $\lambda_{1}=0.091\mum$ to be the Lyman limit and $\lambda_{2}=20\mum$. For a stellar source, $u_{\rm rad}=L_{\star}/(4\pi r^{2}c)$ is the radiation energy density at distance $r$ from a central source of luminosity $L_{\star}$. 

\subsection{Radiation pressure opacity}
%In radiation pressure feedback, usually one is interested in the radiation pressure coefficient per mass unit of dust or of total gas, so-called opacity, assuming dust and gas coupling.

Dust grains have a grain size distribution, which is usually described by a power law,
\begin{align} \label{eq:dnda}
\frac{dn_{j}}{da} = C_{j} n_{\rm H} a^{\alpha},
\end{align} 
where $j$ denotes the grain composition (silicate and graphite), $C_{j}$ is the normalization constant, and $\alpha$ is the power slope. The lower cutoff of the size distribution is taken to be $a_{\rm min}=0.005\mum$ as in \cite{1977ApJ...217..425M}. The upper cutoff of the size distribution, $a_{\rm max}$, is a free parameter in this section.

Integrating over the grain size distribution for all grains in a unit volume, the radiation force on a mass unit is given by
\bea
F_{\rm rad}=\int_{a_{\min}}^{a_{\max}} u_{\rm rad}\bar{Q}_{\rm pr}\pi a^{2} \frac{dn_{gr}}{da}da=u_{\rm rad}\langle\kappa\rangle_{\rm pr,d}\rho_{d},\label{eq:Frad}
\ena
where the radiation pressure coefficient per dust mass is given by
\begin{align} 
\langle\kappa\rangle_{\rm pr,d} = \sum_{j=\rm sil,gra} \int_{a_{\rm min}}^{a_{\rm max}}  \pi a^{2} \bar{Q}_{\rm pr}^{j}(a)\left(\frac{dn_{j}}{da}\right)da,\label{eq:kappad}
\end{align}
and the dust mass density
\bea
\rho_{d}= \sum_{j=\rm sil,gra} \int_{a_{\rm min}}^{a_{\rm max}}  (4/3)\rho_{j}\pi a^{3}\left(\frac{dn_{j}}{da}\right)da,
\ena
where $\rho_{\rm sil}\approx 3.5\g\cm^{-3}$ and $\rho_{\rm gra}\approx 0.2\g\cm^{-3}$  (see e.g., \citealt{2006ApJ...636.1114D}).

Let $f_{d/g}$ be the dust-to-gas mass ratio. Then, the dust radiation pressure coefficient per gas mass, ${\kappa}_{\rm pr}$, is given by
\bea
\kappa_{\rm pr}=\langle\kappa\rangle_{\rm pr,d}f_{d/g}.\label{eq:kappa_pr}
\ena

\subsection{Dependence of Radiation Pressure Opacity on Incident Radiation and Grain Size Distribution}

\begin{figure}
\includegraphics[width=0.5\textwidth]{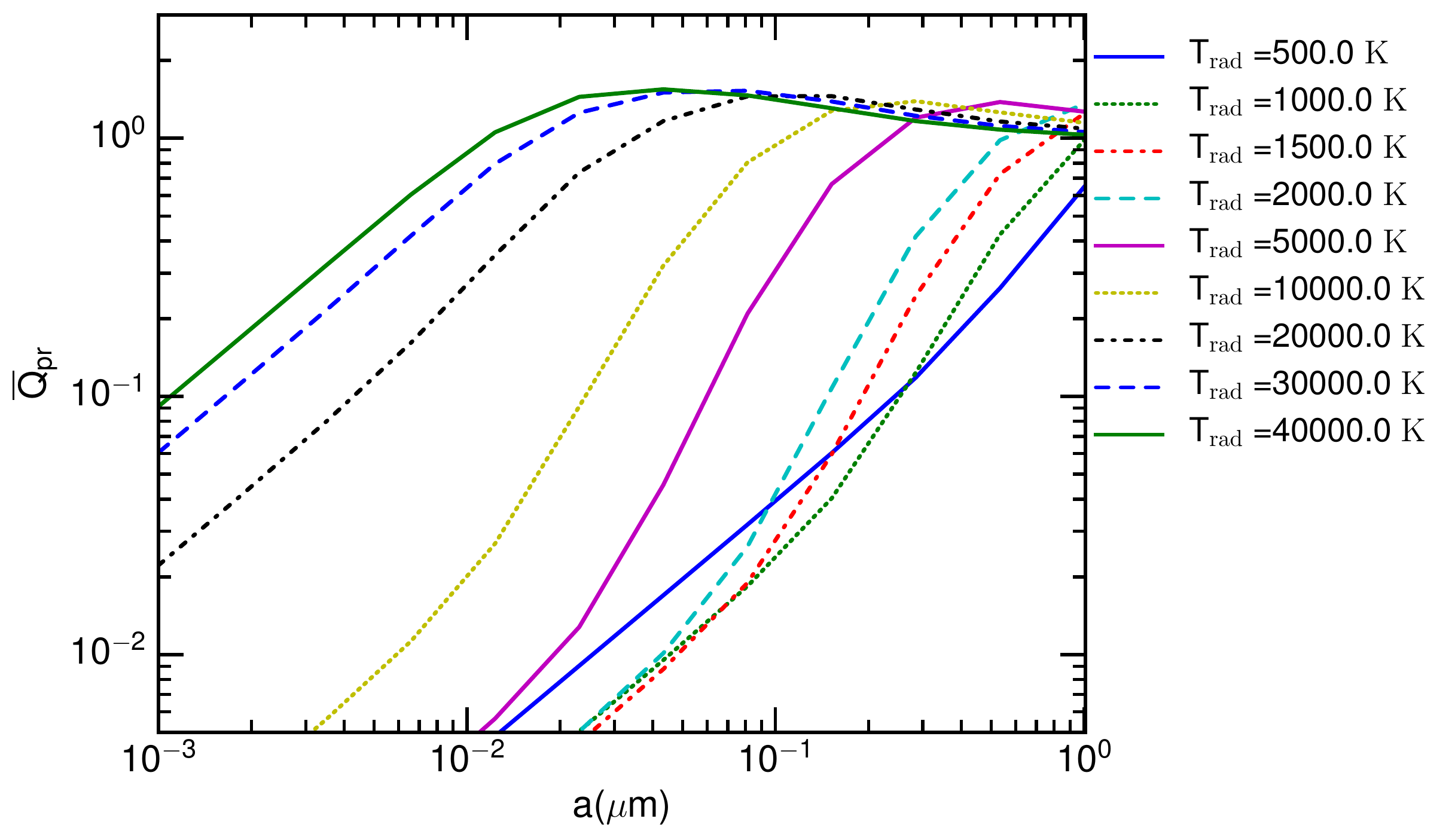}
\caption{Average radiation pressure cross-section efficiency for different temperatures of the black body radiation spectrum, $T_{\rm rad}$, as a function of the grain size for silicates. Rapid decrease of $\bar{Q}_{\rm pr}$ with the grain size is seen for small grains depending on $T_{\rm rad}$.}
\label{fig:Qpr}
\end{figure}

We first calculate the radiation pressure cross-sections for spherical grains using the Mie theory for astronomical silicate and graphite grains (\citealt{1984ApJ...285...89D}). 
We then calculate the radiation pressure cross-section averaged over the different black body radiation spectra characterized by a black body temperature $T_{\rm rad}$, $\bar{Q}_{\rm pr}$, using Equation (\ref{eq:Qpr_avg}). Figure \ref{fig:Qpr} shows $\bar{Q}_{\rm pr}$ as a function of the grain size for different radiation spectra. For low temperatures of $T_{\rm rad}\lesssim 2000\K$, $\bar{Q}_{\rm pr}$ decreases rapidly when the grain size decreases from $a=1\mum$ because the peak wavelength, $\lambda_{\max}= 2898\mum \K/T_{\rm rad}\sim 3(10^{3}\K/T_{\rm rad})\mum$, is much larger than the grain size. For hot stars of $T_{\rm rad}>2\times 10^{4}\K$, $\bar{Q}_{\rm pr}$ first increases slowly to its maximum and then decreases when the grain size becomes small enough ($\sim 0.01-0.1\mum$) such that $\lambda_{\max}/a>1$.
 
\begin{figure*}
\includegraphics[width=0.5\textwidth]{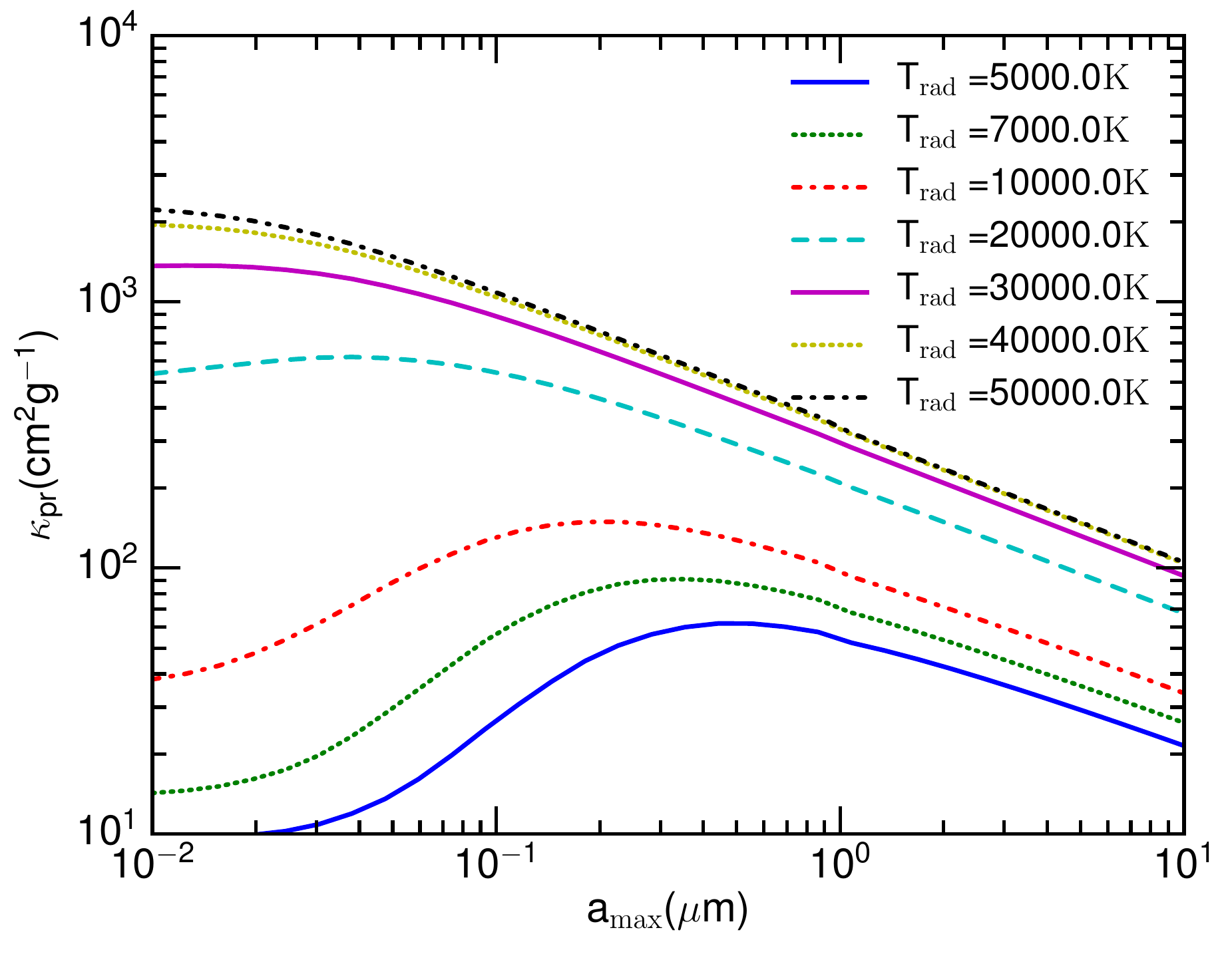}
\includegraphics[width=0.5\textwidth]{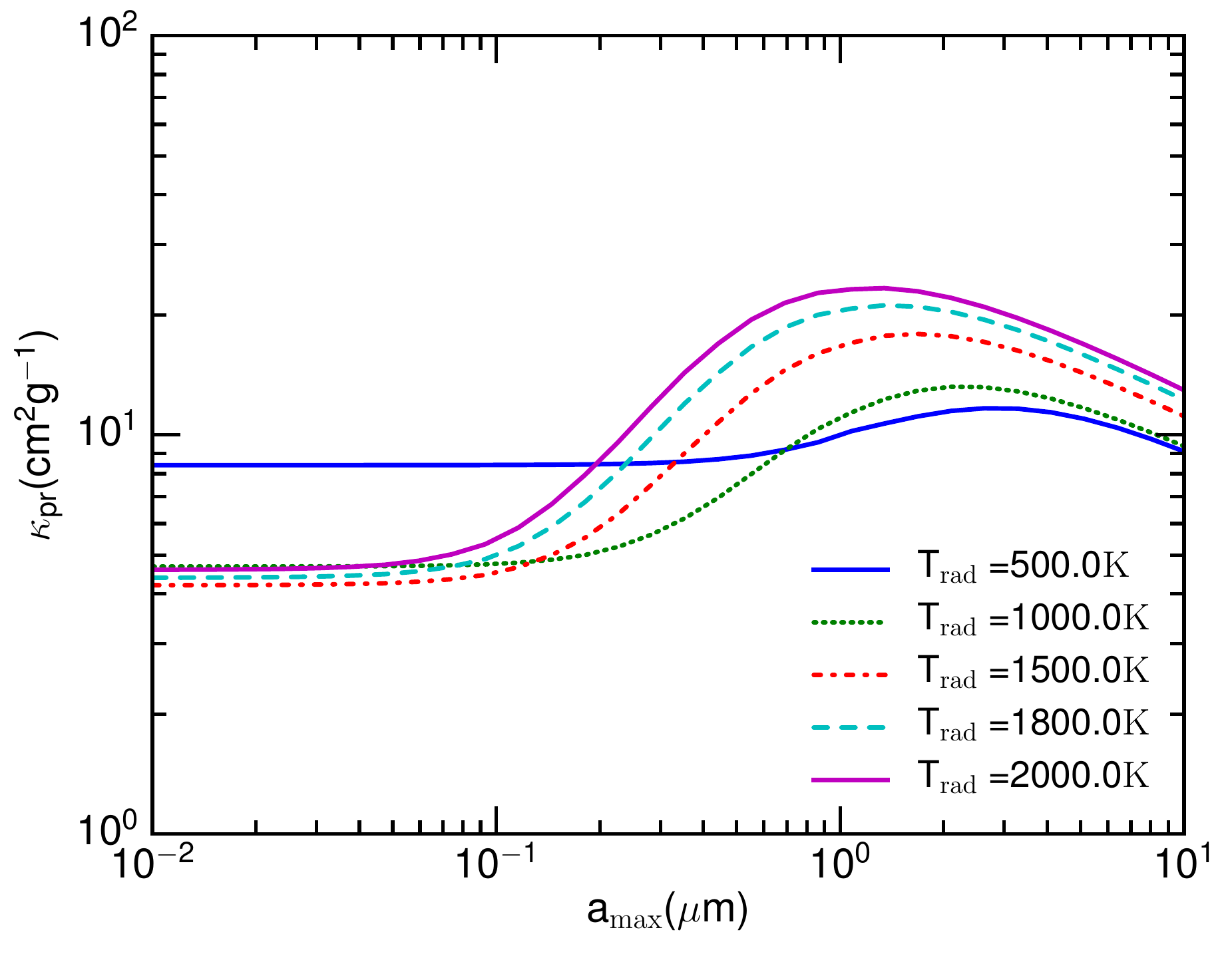}
\includegraphics[width=0.5\textwidth]{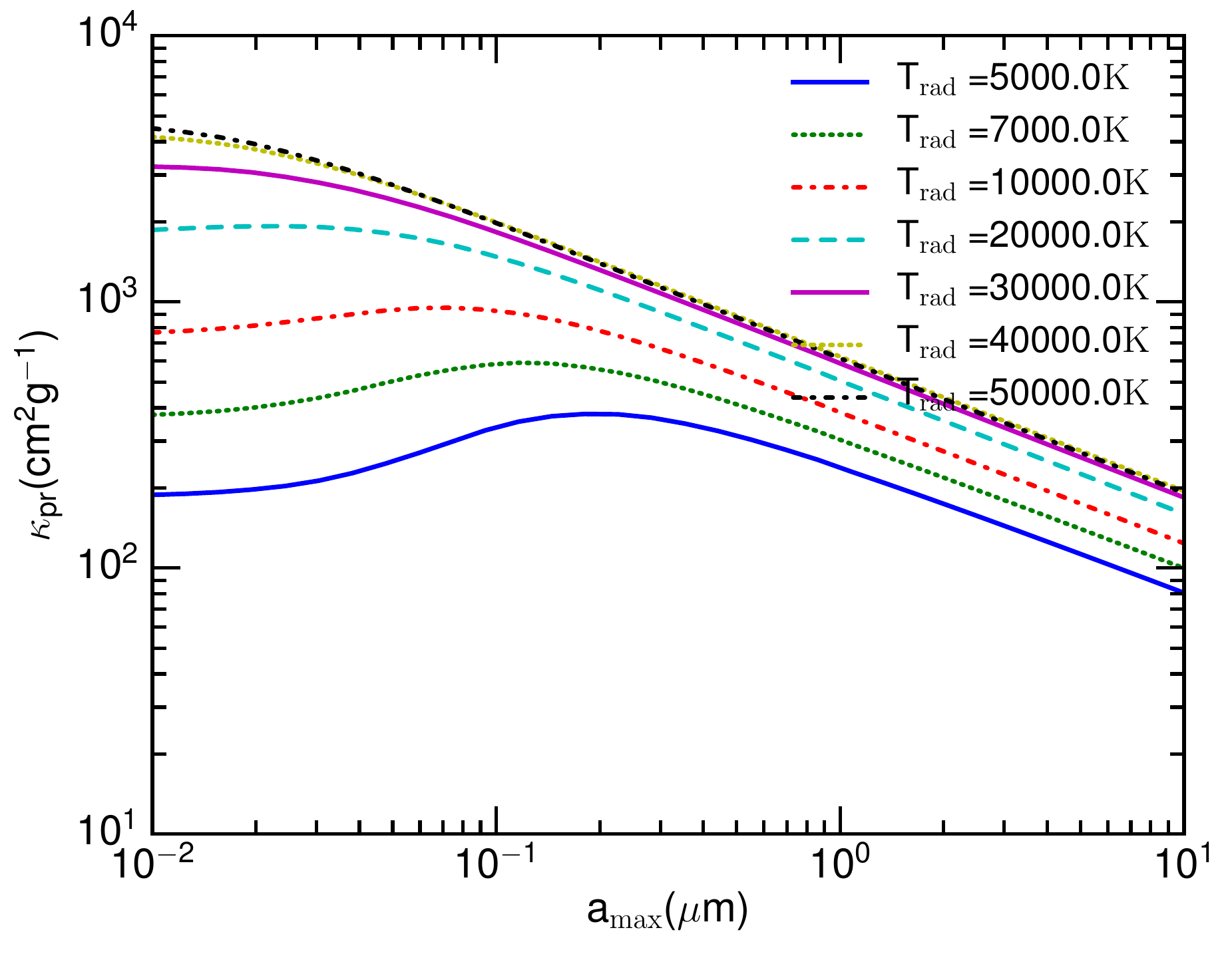}
\includegraphics[width=0.5\textwidth]{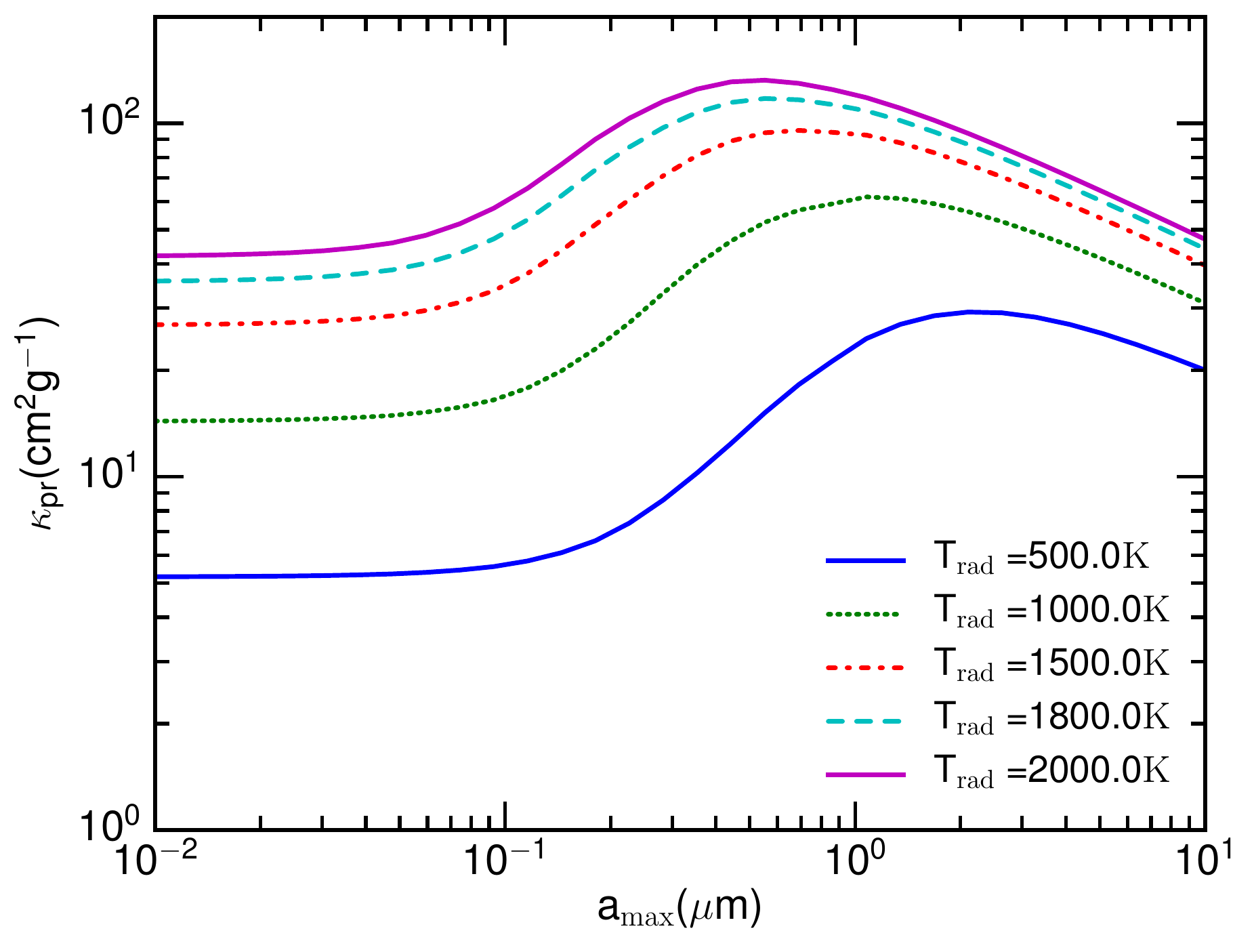}
\caption{Effect of the maximum grain size $a_{\max}$ on radiation pressure opacity for different black body radiation spectra from the central star (left panel) and hot dust shell (right panel), assuming silicate material (upper panels) and graphite (lower panels). The opacity $\kappa_{\rm pr}$ decreases rapidly when $a_{\max}$ decreases and becomes flat for $a\lesssim 0.1\mum$.}
\label{fig:kappa_pr}
\end{figure*}

To study the effects of the grain size distribution on radiation pressure, we calculate the radiation pressure opacity for the different values of $a_{\max}$ and a constant slope $\alpha=-3.5$ using Equations (\ref{eq:kappad}) and (\ref{eq:kappa_pr}).

Figure \ref{fig:kappa_pr} (upper) shows the effect of the maximum size on the radiation pressure opacity averaged over the stellar radiation and infrared dust emission field. For hot stars of high stellar temperatures of $T_{\rm rad}>20000K$, the radiation opacity increases when $a_{\max}$ decreases from $1\mum$ to $0.01\mum$ because the radiation spectrum from hot stars is mostly in the UV. For low source temperatures (lower panel), the opacity $\kappa_{\rm pr}$ decreases rapidly when $a_{\max}$ decreases to $\lesssim 0.1\mum$ because these cool sources emit radiation mostly in optical and near-infrared (NIR). For the typical of the hot dust shell of $T_{\rm rad}\sim 1000\K$ adopted by WC87, the opacity decreases by a factor of $3$ when the maximum size is reduced to $a_{\max}=0.1\mum$.

The results shown in Figure \ref{fig:kappa_pr} reveal that the maximum grain size is an important factor of dust radiation pressure coefficient. Therefore, an accurate understanding of how $a_{\max}$ changes in the protostellar envelope is critically important.
%The effect of $a_{\max}$ on dust mass absorption coefficient is studied in \cite{2019arXiv190905015Y}, which shows the importance of $a_{\max}$ for the different environments. 
In the following section, we will show that the maximum size $a_{\max}$ decreases with increasing radiation intensity, which make the opacity to change with the distance to the radiation source accordingly.

%{\bf We stress that the difference in $\kappa_{\rm pr}$ by a factor of 5 corresponds to the difference in the radiation pressure force by a factor of $e^{5}\sim 150$ times, which is substantial.}

%The opacity calculated in \cite{2010ApJ...722.1556K} using Laor and Draine (1993) for silicate grains from $a=0.005-10\mum$ is $\kappa_{\rm pr}\sim 5cm^{-2}/g$ for $T_{d}\sim 1000\K$.

\section{Radiative Torques and Grain Rotational Disruption}\label{sec:RATs}
Dust grains of irregular shape irradiated by an anisotropic radiation experience radiative torques (\citealt{1976Ap&SS..43..291D}; \citealt{1996ApJ...470..551D}; \citealt{2004ApJ...614..781A}). The magnitude of RATs is defined as
\bea
{\Gamma}_{\lambda}=\pi a^{2}
\gamma u_{\lambda} \left(\frac{\lambda}{2\pi}\right){Q}_{\Gamma},\label{eq:GammaRAT}
\ena
where $\gamma$ is the anisotropy degree of the radiation field, ${Q}_{\Gamma}$ is the RAT efficiency (\citealt{1996ApJ...470..551D}; \citealt{2007MNRAS.378..910L}).\footnote{Formally, $a$ here is the effective size of the grain which is defined as the radius of the sphere with the same volume as the irregular grain, but for simplicity, we take $a$ without significant uncertainty (within a order of unity).}

The magnitude of RAT efficiency can be approximated by a power-law
\bea
Q_{\Gamma}\approx \alpha\left(\frac{{\lambda}}{a}\right)^{-\eta}\label{eq:QAMO}
\ena
for $\lambda/a\gtrsim 0.1$, where $\alpha$ and $\eta$ are the constants that depend on the grain size, shape, and optical constants. Numerical calculations of RATs for several shapes of different optical constants in \cite{2007MNRAS.378..910L} find the slight difference in RATs among the realization. They adopted the coefficients $\alpha=0.4,\eta=0$ for $a_{\rm trans}<a<\lambda/0.1$, and $\alpha=2.33,\eta=3$ for $a<a_{\rm trans}$ where $a_{\rm trans}=\lambda/1.8$ denotes the transition size at which the RAT efficiency slope changes. Thus, the maximum RAT efficiency is $Q_{\Gamma,\max}=\alpha$. 

The radiative torque averaged over the incident radiation spectrum is defined as
\bea
\overline{\Gamma}_{\rm RAT}&=&\int {\Gamma}_{\lambda}d\lambda=\pi a^{2}
\gamma u_{\rad} \left(\frac{\overline{\lambda}}{2\pi}\right)\overline{Q}_{\Gamma},\label{eq:GammaRAT_num}
\ena
where the average radiative torque efficiency over the radiation spectrum is defined as
\bea
\overline{Q}_{\Gamma} = \frac{\int_{0}^{\infty} \lambda Q_{\Gamma}u_{\lambda} d\lambda}{\int_{0}^{\infty} \lambda u_{\lambda} d\lambda}=\frac{\int_{0}^{\infty} \lambda Q_{\Gamma}u_{\lambda} d\lambda}{\bar{\lambda}u_{\rm rad}},\label{eq:Qavg_num}
\ena
where the integrals are taken over the entire radiation spectrum.

For a radiation spectrum of black body temperature $T_{\rm rad}$, the mean wavelength of the stellar radiation field is given by
\bea
\bar{\lambda}(T_{\rad})&&=\frac{\int_{0}^{\infty} \lambda B_{\lambda}(T_{\rm rad})d\lambda}{\int _{0}^{\infty}B_{\lambda}(T_{\rm rad})d\lambda}\nonumber\\
&=&\left(\frac{2\pi k^{3}\Gamma(3)\zeta(3)}{\sigma ch^{2}}\right)\frac{1}{T_{\rad}}\simeq \frac{0.53\cm \K}{T_{\rm rad}},\label{eq:wavemean_star}
\ena
where $\Gamma$ and $\zeta$ are the Gamma and Riemann functions, and we have used the integral formula $\int_{0}^{\infty} x^{s-1}dx/(e^{x}-1)=\Gamma(s)\zeta(s)$ for $s>1$.

For small grains of $a<\bar{\lambda}/1.8$, plugging $Q_{\Gamma}$ from Equation (\ref{eq:QAMO}) and $u_{\lambda}\propto B_{\lambda}(T_{\rad})$ into Equation (\ref{eq:Qavg_num}), one obtains the following after taking the integral,
\bea
\overline{Q}_{\Gamma}&=&\frac{2\pi \alpha k^{\eta+3}}{\sigma h^{\eta+2}c^{\eta+1}}\left(\frac{\zeta(3)\Gamma(3)2\pi k^{3}}{\sigma ch^{2}} \right)^{\eta-1}\Gamma(\eta+3)\zeta(\eta+3)\nonumber\\
&&\times\left(\frac{\bar{\lambda}}{a}\right)^{-\eta}.\label{eq:Qmean_star}
\ena

Plugging the RAT parameters of $\alpha=2.33$ and $\eta=3$ into Equation (\ref{eq:Qmean_star}), one obtains the average RAT efficiency for a stellar radiation field
\bea
\overline{Q}_{\Gamma}\simeq %5.98\left(\frac{\bar{\lambda}}{a}\right)^{-3}.\label{eq:Qavg_star}
6\left(\frac{\bar{\lambda}}{a}\right)^{-3}.\label{eq:Qavg_star}
\ena 
Because the average RAT efficiency cannot exceed its maximum RAT efficiency, $Q_{\Gamma,\max}$, the above equation is only valid for grains of size $a\lesssim (Q_{\Gamma,\max}/6)^{1/3}\bar{\lambda}= \bar{\lambda}/2.5$. Large grains of $a>\bar{\lambda}/2.5$ then have $\bar{Q}_{\Gamma}=Q_{\Gamma,\max}=0.4$. Let $a_{\rm trans,\star}\equiv\bar{\lambda}/2.5$ be the transition size of the RAT averaged over the stellar radiation spectrum.

\subsection{Grain Rotation and Rotational Disruption by Radiative Torques}\label{sec:disr}
For radiation sources with stable luminosity considered in this paper, radiative torque, $\overline{\Gamma}_{\rm RAT}$, is constant, so that the grain angular velocity is steadily increased over time. The equilibrium angular velocity can be achieved when the spin-up rate by RATs is equal to the damping rate (see \citealt{2007MNRAS.378..910L}; \citealt{2009ApJ...695.1457H}; \citealt{2014MNRAS.438..680H}):
\bea
\omega_{\rm RAT}=\frac{\overline{\Gamma}_{\rm RAT}\tau_{\rm damp}}{I},\label{eq:omega_RAT0}
\ena
where $\tau_{\rm damp}$ is the rotational damping time (see Eq. \ref{eq:taudamp}), which is induced by gas collisions and IR emission (see Appendix \ref{sec:appendix}).

For the gas with hydrogen density $n_{\H}$ and temperature $T_{\gas}$, plugging $\tau_{\rm damp}$ and $\overline{\Gamma}_{\rm RAT}$ with $Q_{\rm RAT}$ from Equation (\ref{eq:Qavg_star}) into Equation (\ref{eq:omega_RAT0}), one obtains
\bea
%\omega_{\rm RAT}&\simeq &3.2\times 10^{7}\gamma_{-1} a_{-5}^{0.7}\bar{\lambda}_{0.5}^{-1.7}\nonumber\\
\omega_{\rm RAT}&=& \frac{3\gamma u_{\rm rad}a\bar{\lambda}^{-2}}{1.6n_{\rm H}\sqrt{2\pi m_{\rm H}kT_{\rm gas}}}\left(\frac{1}{1+F_{\rm IR}}\right)\nonumber\\
&\simeq &9.4\times 10^{5} a_{-5}\left(\frac{\bar{\lambda}}{1.2\mum}\right)^{-2}
\left(\frac{\gamma_{-1}U}{n_{3}T_{1}^{1/2}}\right)\nonumber\\
&&\times\left(\frac{1}{1+F_{\rm IR}}\right)\rad\s^{-1},\label{eq:omega_RAT1}
\ena
for grains with $a\lesssim a_{\rm trans, \star}$, and
\bea
\omega_{\rm RAT}&=&\frac{1.5\gamma u_{\rm rad}\bar{\lambda}a^{-2}}{16n_{\rm H}\sqrt{2\pi m_{\rm H}kT_{\rm gas}}}\left(\frac{1}{1+F_{\rm IR}}\right)\nonumber\\
&\simeq& 8.1\times 10^{7}a_{-5}^{-2}\left(\frac{\bar{\lambda}}{1.2\mum}\right) \left(\frac{\gamma_{-1}U}{n_{3}T_{1}^{1/2}}\right)\nonumber\\
&&\times\left(\frac{1}{1+F_{\rm IR}}\right)\rad\s^{-1},\label{eq:omega_RAT2}
\ena
for grains with $a> a_{\rm trans, \star}$. Here $a_{-5}=a/10^{-5}\cm$, $n_{3}=n_{\H}/10^{3}\cm^{-3}$, $T_{1}=T_{\gas}/10\K$, $F_{\rm IR}$ is the IR damping coefficient (see Eq. \ref{eq:FIR}), and $\gamma_{-1}=\gamma/0.1$ is the anisotropy of radiation field relative to the typical anisotropy of the diffuse interstellar radiation field of $\gamma=0.1$ (e.g., \citealt{1996ApJ...470..551D}). The stellar radiation field has $\gamma=1$.

A spherical dust grain of radius $a$ rotating at velocity $\omega$ develops an average tensile stress due to centrifugal force which scales as (see \citealt{Hoang:2019da})
\bea
S=\frac{\rho a^{2} \omega^{2}}{4},\label{eq:Stress}
\ena 
where $\rho$ is the dust mass density.

When the rotation rate is sufficiently high such as the tensile stress exceeds the maximum limit (i.e., tensile strength), $S_{\rm max}$, the grain is disrupted. The critical rotational velocity is given by $S=S_{\rm max}$:
\bea
\omega_{\rm disr}&=&\frac{2}{a}\left(\frac{S_{\max}}{\rho} \right)^{1/2}\nonumber\\
&\simeq& \frac{3.65\times 10^{8}}{a_{-5}}S_{\max,7}^{1/2}\hat{\rho}^{-1/2}~\rad\s^{-1},\label{eq:omega_cri}
\ena
where $\hat{\rho}=\rho/(3\g\cm^{-3})$, and $S_{\max,7}=S_{\max}/10^{7} \erg\cm^{-3}$ (\citealt{Hoang:2019da}).
 
The tensile strength of interstellar dust depends on grain structure, which is uncertain (\citealt{1990ARA&A..28...37M}). Compact grains have large tensile strength of $S_{\rm max}\gtrsim 10^{9}\erg\cm^{-3}$, whereas composite/fluffy grains have a much lower tensile strength \citep{2019ApJ...876...13H}. Large interstellar grains (radius $a>0.1\mum$) are expected to have a composite structure (\citealt{1989ApJ...341..808M}; \citealt{Draine:2020ua}) as a result of coagulation process in molecular clouds or in the interstellar medium (ISM). Numerical simulations for porous grain aggregates from \cite{2019ApJ...874..159T} find that the tensile strength decreases with increasing the monomer radius and can be fitted with an analytical formula (see \citealt{2020MNRAS.496.1667K} for more details)
\bea
S_{\max} &\simeq& 9.51\times 10^{4} \left(\frac{\gamma_{\rm sf}}{100 \erg\cm^{-2}}\right) \nonumber\\
&\times&\left(\frac{r_{0}}{0.1\mum}\right)^{-1}\left(\frac{\phi}{0.1}\right)^{1.8} \erg\cm^{-3},\label{eq:Smax}
\ena
where $\gamma_{\rm sf}$ is the surface energy per unit area of the material, $r_{0}$ is the monomer radius, and $\phi$ is the volume filling factor of monomers. For large grains ($a>0.1\mum$) made of ice-mantle monomers of radius $r_{0}=0.1\mum$ and $\phi=0.1$, Equation (\ref{eq:Smax}) implies $S_{\max}\approx 10^{5}\erg\cm^{-3}$, assuming the surface energy of $\gamma_{\rm sf}=0.1 J m^{-2}$ for ice mantles in contact. We note that according to the dust evolution, grains in dense cores are expected to be porous aggregates due to the coagulation of ice mantles grains because ice mantles develop at $A_{V}\sim 3$. Therefore, the typical value of $S_{\max}$ for dust in the massive prestellar cores is $S_{\max}\sim 10^{5} \erg\cm^{-3}$.

Comparing Equations (\ref{eq:omega_RAT1}) and (\ref{eq:omega_cri}), one can obtain the disruption grain size:
\bea
a_{\rm disr}&=&\left(\frac{3.2n_{\rm H}\sqrt{2\pi m_{\rm H}kT_{\rm gas}}}{3\gamma u_{\rm rad}\bar{\lambda}^{-2}}\right)^{1/2}\left(\frac{S_{\rm max}}{\rho}\right)^{1/4}(1+F_{\rm IR})^{1/2}\nonumber\\
&\simeq& 1.96 \left(\frac{\gamma_{-1}U}{n_{3}T_{1}^{1/2}}\right)^{-1/2}\left(\frac{\bar{\lambda}}{1.2\mum}\right) \hat{\rho}^{-1/4}S_{\max,7}^{1/4}\nonumber\\
&&\times (1+F_{\rm IR})^{1/2}\mum,
\label{eq:adisr_ana}
\ena
which depends on the local gas properties, radiation field, and the grain tensile strength. 

Due to the decrease of the rotation rate for $a>a_{\rm trans,\star}$ (see Eq. \ref{eq:omega_RAT2}), the rotational disruption occurs only if $a_{\rm disr}<a_{\rm trans, \star}$. In this case, there exists a maximum size of grains that can still be disrupted by centrifugal stress (\citealt{2020ApJ...891...38H}),
\bea
a_{\rm disr,max}&=&\frac{\gamma u_{\rm rad}\bar{\lambda}}{12n_{\rm H}\sqrt{2\pi m_{\rm H}kT_{\rm gas}}}\left(\frac{S_{\rm max}}{\rho}\right)^{-1/2}(1+F_{\rm IR})^{-1}\nonumber\\
&\simeq& 
0.04\left(\frac{\gamma_{-1} U}{n_{3}T_{1}^{1/2}}\right)\left(\frac{\bar{\lambda}}{1.2\mum}\right)\hat{\rho}^{1/2}S_{\max,7}^{-1/2}\nonumber\\
&\times&(1+F_{\rm IR})^{-1}\mum.\label{eq:adisr_up}
\ena 

In general, due to dependence of $F_{\rm IR}$ on the grain size $a$, one only obtain analytical results for $a_{\rm disr}$ when $F_{\rm IR}\ll 1$. In general, we first calculate numerically $\omega_{\rm RAT}$ using Equation (\ref{eq:omega_RAT0}) and compare it with $\omega_{\rm disr}$ to find $a_{\rm disr}$ numerically, which will be referred to as numerical results.

\subsection{Rotational Disruption vs. Radiation Pressure Acceleration} 
Under the intense radiation field, grains are also known to be accelerated by radiation force. To see whether grain acceleration can be more effecient than rotational disruption, we now compare the characteristic timescales of these two processes.

The characteristic timescale for rotational disruption of a grain of size $a$ can be estimated as (\citealt{Hoang:2019da}):
\bea
t_{\rm disr}&=&\frac{I\omega_{\rm disr}}{\Gamma_{\rm RAT}}=\frac{I\omega_{\rm disr}}{\pi a^{2}u_{\rm rad}(\bar{\lambda}/2\pi)\overline{Q}_{\Gamma}},\nonumber\\
&=&\frac{32\pi a^{2}(\rho S_{\max})^{1/2}}{15u_{\rm rad}\bar{\lambda}\overline{Q}_{\Gamma}}\nonumber\\
&&\simeq 101a_{-5}^{2}(\hat{\rho}S_{\max,7})^{1/2}\left(\frac{r}{100\AU}\right)^{2}\left(\frac{0.2\mum}{\bar{\lambda}\bar{Q}_{\Gamma}}\right)\s.~~~~~\label{eq:tdisr}
\ena

The characteristic timescale to accelerate the grain at rest to a velocity $v$ by radiation pressure is estimated as,
\bea
t_{\rm acc}&&=\frac{m_{gr}v}{F_{\rm rad}(a)}=\frac{4\rho a v}{3\bar{Q}_{\rm pr}u_{\rm rad}}\nonumber\\
&&\simeq
8815a_{-5}v_{1}\hat{\rho}\left(\frac{r}{100\AU}\right)^{2}\left(\frac{1.0}{\bar{Q}_{\rm pr}L_{6}}\right)\s,~~~~~
\label{eq:tacc}
\ena
where $v_{1}=v/(10\km\s^{-1})$.

The ratio of the disruption time to acceleration time is then equal to
\bea
\frac{t_{\rm disr}}{t_{\rm acc}}&=&\frac{8\pi a Q_{\rm pr}(S_{\max}/\rho)^{1/2}}{5 v\bar{\lambda}\overline{Q}},\nonumber\\
&\simeq& 0.01a_{-5}\left(\frac{\bar{Q}_{\rm pr}}{1.0}\right)\frac{(S_{\rm max,7}/\hat{\rho})^{1/2}}{v_{1}(\bar{\lambda}/0.2\mum)(\overline{Q}_{\Gamma}/0.4)},\label{eq:tdisr_acc}
\ena
where the radiation pressure and torque coefficients are normalized over their typical values of $\bar{Q}_{\rm pr}= 1$ and $\overline{Q}_{\Gamma}= 0.4$. 

Equation (\ref{eq:tdisr_acc}) implies that the rotational disruption occurs on a much shorter timescale compared to acceleration by radiation pressure for $S_{\max}<10^{11}\erg\cm^{-3}$. Therefore, essentially, grains of both porous and compact structures can be disrupted before being accelerated by radiation pressure. Therefore, the radiation pressure opacity would be determined by the dust size distribution constrained by RATD instead of the original size distribution of dust in prestellar cores on the standard interstellar dust model. We will quantify the effect of RATD on the radiation opacity in the next section. The special case is in an extremely high density where grain rotational damping by gas collisions is faster than disruption. For this case, grains cannot be disrupted, but dust grains could be slowly accelerated and coupled to the gas such as in stellar winds.

Let's compare the disruption time the dynamical times of the star formation process. The characteristic timescale for a molecular core to collapse to form a zero main-sequence star is described by the Kelvin-Helmholtz timescale,
\bea
t_{\rm KH}&&=\frac{GM_{\star}^{2}}{R_{\star}L}\nonumber\\
&&\simeq 3140 \left(\frac{M_{\star}}{100M_{\odot}}\right)^{2}\left(\frac{10R_{\odot}}{R_{\star}}\right)\left(\frac{10^{6}L_{\odot}}{L}\right)\yr,
\ena
which is much longer than the disruption and acceleration time (see Eq.\ref{eq:tacc}). The free-fall time of gravitational collapse is,
\bea
t_{\rm ff}=\left(\frac{3\pi}{32 G\rho_{\gas}}\right)^{1/2}\simeq 1.63\times 10^{4}n_{7}^{-1/2}\yr,
\ena
where $n_{7}=n_{\H}/10^{7}\cm^{-3}$ is normalized over the density of the central core with $\rho_{\gas}=\mu n_{\H}m_{\H}$ with $\mu$ being the molecular weight of the gas. This is also much longer than the acceleration and disruption times.

%This ratio would determin the upper limit of the grain velocity achieved by radiation pressure for large grains. Small grains could be accelerated then.

\section{Model of a Massive Protostellar Core}\label{sec:protostar}

\subsection{Density profile}
Let $\dot{M}$ be the accretion rate of the matter onto the central massive protostar of mass $M_{\star}$. The accretion rate is determined by the initial condition. For free-fall accretion, the inflow velocity at radius $r$ is $v(r)=(2GM_{\star}/r)^{1/2}$. The gas density is then given by
\bea
n_{\H}(r)&=&\frac{\dot{M}X(\H)}{4\pi r^{2}m_{\H}v}\nonumber\\
&\simeq& 2.2\times 10^{8}\frac{X(\H)}{0.7}\left(\frac{M_{\star}}{100M_{\odot}}\right)^{-1/2}\left(\frac{\dot{M}}{10^{-3}M_{\odot}\yr^{-1}}\right)\nonumber\\
&&\times\left(\frac{r}{100\AU}\right)^{-3/2}\cm^{-3}.\label{eq:nH}
\ena
where $X(\H)$ is the mass fraction of hydrogen mass in the infalling gas. The density profile depends on the stellar mass and accretion rate.

\subsection{Radiation field of the central massive star}
For massive stars, nuclear fusion already started when the accretion process is still ongoing, which is different from the low-mass protostar. The total luminosity from the central protostar is given by
\bea
L_{\rm tot}=\int_{0}^{\infty} L_{\nu}d\nu = L_{\star}+L_{\rm acc},
\ena
where $L_{\star}$ is the stellar luminosity produced by the nuclear burning core, and $L_{\rm acc}$ is the luminosity induced by accretion shock at the star surface.
The latter is equal to the rate of gravitational energy released due to collapse,
\bea
L_{\rm acc}&=&\frac{GM_{\star}\dot{M}}{R_{\star}}\simeq3.1\times10^{5}L_{\odot}\left(\frac{M_{\star}}{100M_{\odot}}\right)\nonumber\\
&&\times \left(\frac{\dot{M}}{10^{-3}M_{\odot}\yr^{-1}}\right)\left(\frac{R_{\star}}{10R_{\odot}}\right),\label{eq:Lacc}
\ena
where $M_{\star}$ is the core mass and $R_{\star}$ core radius. 

Following \cite{1980A&A....92..101M}, one has $L_{\star}=10^{6}(M/100M_{\odot})^{\alpha}$ with $\alpha \sim 1.5$ for $M\sim 100 M_{\odot}$. Thus, the stellar luminosity dominates over the accretion luminosity for very massive stars.

The central star is assumed to radiate as a black body from photosphere of temperature $T_{\rm ph}$ (e.g., \citealt{1986ApJ...310..207W}; \citealt{1987ApJ...319..850W}). Thus, the stellar temperature is related to the stellar luminosity as 
\bea
L_{\rm tot}=4\pi R_{\rm ph}^{2}\sigma T_{\rm ph}^{4},
\ena
where $R_{\rm ph}$ is the stellar photosphere radius. %For $R_{\rm ph}\sim 10^{12}$ and $T_{\rm ph}\sim 20,000K$, the luminosity is $L_{\star}\sim 3\times 10^{6}L_{\odot}$.

%We adopt the values of $T_{\rm ph}\sim 30000\K$ for $\dot{M}=10^{-3}-10^{-4}$ from \cite{1986ApJ...310..207W}. We also consider two values of stellar masses of $M_{\star}=60$ and $100M_{\odot}$ as in WC87.

\begin{figure}
\includegraphics[width=0.5\textwidth]{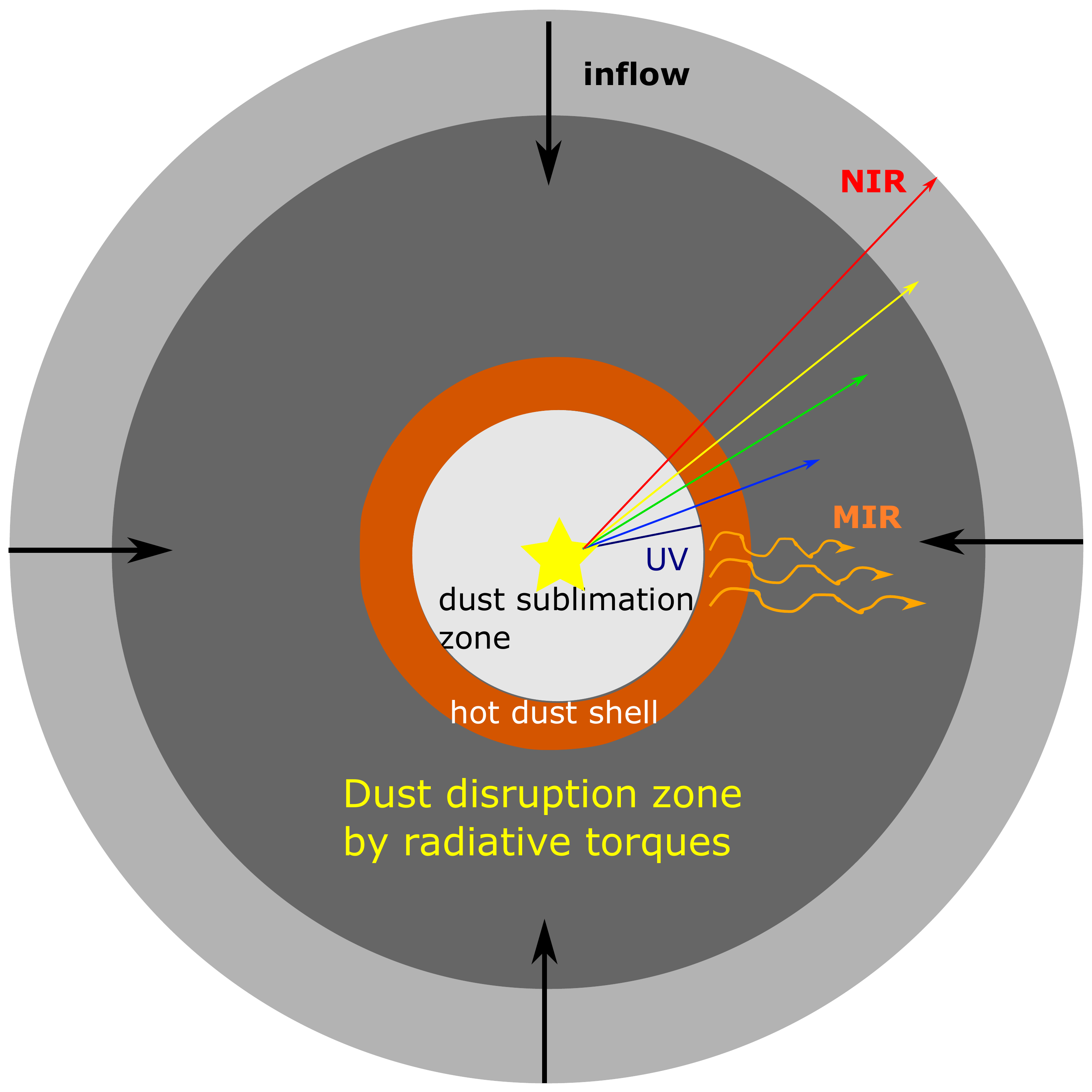}
\caption{Schematic illustration of a protostellar envelope, including dust-free sublimation zone, first dust absorption zone (also disruption zone), and outer stellar-shielded envelope. Disruption of grains in the first absorption zone changes the radiation spectrum irradiating the outer zone as well as radiation acceleration of the first zone.}
\label{fig:massivestar}
\end{figure}

\subsubsection{Inner envelope: irradiated by direct stellar radiation}
Dust near the central star is sublimated due to heating by intense stellar radiation. The radius of the dust sublimation front is given by setting the grain temperature $T_{d}$ to be equal to the sublimation threshold, $T_{\rm sub}$, which yields (see \citealt{2015ApJ...806..255H})
\bea
r_{\rm sub}\simeq 155.3\left(\frac{L_{\rm tot}}{10^{6}L_{\odot}}\right)^{1/2}\left(\frac{T_{\rm sub}}{1500\K}\right)^{-5.6/2}\AU.\label{eq:rsub}
\ena

Due to the extinction by intervening dust, the radiation strength of the stellar radiation field at radial distance $r$ from the central star is given by
\bea
U_{\star}(r)=\frac{\int_{0}^{\infty} u_{\lambda}(T_{\star})e^{-\tau(\lambda)}d\lambda}{u_{\rm ISRF}},\label{eq:U_star_red}
\ena
where $u_{\lambda}(T_{\star})=L_{\lambda}/(4\pi r^{2}c)$ is the spectral energy density in the absence of dust extinction, $\tau_{\lambda}$ is the optical depth of intervening dust.

The mean wavelength of the reddened stellar spectrum is given by Equation (\ref{eq:wavemean_star}) with $u_{\lambda}(T_{\star})\rightarrow u_{\lambda}(T_{\star})e^{-\tau_{\lambda}}$.

%Practically, for numerical integration, we take the lower limit of $\lambda_{1}=0.091\mum$ (Lyman limit) and the upper limit of $\lambda_{2}=20\mum$ for Equations (\ref{eq:U_star_red}) and (\ref{eq:wavemean_star_red}).

For massive stars, due to dominance of UV photons, the stellar radiation is mostly absorbed by a thin shell of visual extinction of $\sim 1$ beyond the sublimation front (see Figure \ref{fig:massivestar}).

\subsubsection{Outer envelope: irradiated by thermal emission from hot dust}
Dust grains in the hot dust shell just beyond the sublimation front are irradiated by direct stellar radiation and thermal emission from the hot dust shell (see Figure \ref{fig:massivestar}). Assuming that the hot dust shell emits as a black body, the total luminosity emitted by the hot shell is equal to the bolometric luminosity $L_{\rm tot}$, which has a specific luminosity of
\bea
L_{\rm shell,\nu}=4\pi R_{\rm shell}^{2}F_{\nu}=4\pi R_{\rm shell}^{2}\pi B_{\lambda}(T_{\rm shell}),
\ena
where $R_{\rm shell}$ is the radius of the hot dust shell, and $F_{\nu}=\pi B_{\nu}(T_{\rm shell})$ is the spectral emergent flux from the thin shell.

The thin hot dust shell is assumed to have temperature of $T=T_{\rm shell}$, given by 
\bea
L=\int L_{\rm shell,\nu}d\nu =4\pi R_{\rm shell}^{2} \sigma T_{\rm shell}^{4}.\label{eq:Ltd}
\ena

The spectral energy density of thermal dust emission at distance $r$ is given by
\bea
u_{\lambda,\rm shell}=\frac{4\pi R_{\rm shell}^{2}\pi B_{\lambda}(T_{\rm shell})}{4\pi r^{2}c}.\label{eq:u_td}
\ena

The mean wavelength of the reddened stellar spectrum is given by Equation (\ref{eq:wavemean_star}) with $u_{\lambda}(T_{\star})\rightarrow u_{\lambda,\rm shell}e^{-\tau_{\lambda}}$. The mean wavelength of the hot dust emission is $\bar{\lambda}_{\rm shell}= 0.53\rm \K/T_{\rm shell}\sim 5.3\mum (1000\K/T_{\rm shell})$. Thus, NIR-MIR emission from hot dust is important for disruption of grains in the outer layer, which are important for large grains only of $a>\bar{\lambda}_{\rm shell}/2\sim 2.7\mum$. Smaller grains can also be disrupted if the luminosity is sufficiently large.

\begin{table*}
\begin{center}
\caption{Best-fit parameters of the radiation strength and mean wavelength to numerical calculations for different stellar temperatures.}\label{tab:coefficient}
\begin{tabular}{l l l l l} \hline\hline
{$T_{\star}(\K)$} & $c_{1}$ & $c_{2}$  & $c_{3}$  & $c_{4}$ \cr
\hline\\
%updated values
1000 & 0.037 $\pm$ 0.0025 & 1.18 $\pm$ 0.021 & 0.05 $\pm$ 0.005 & 0.75 $\pm$ 0.027\cr
1500 & 0.067 $\pm$ 0.0019 & 1.24 $\pm$ 0.009 & 0.08 $\pm$ 0.006 & 0.75 $\pm$ 0.021\cr
2000 & 0.13 $\pm$ 0.0047 & 1.21 $\pm$ 0.014 & 0.13 $\pm$ 0.007 & 0.71 $\pm$ 0.015\cr
2500 & 0.19 $\pm$ 0.008 & 1.24 $\pm$ 0.019 & 0.15 $\pm$ 0.007 & 0.74 $\pm$ 0.014\cr

\hline
10000 & 1.68 $\pm$ 0.061 & 1.62 $\pm$ 0.052 & 0.86 $\pm$ 0.036 & 0.65 $\pm$ 0.012\cr
20000 & 3.68 $\pm$ 0.028& 2.12 $\pm$ 0.024 & 1.91 $\pm$ 0.081 & 0.62 $\pm$ 0.012 \cr
30000 & 4.94 $\pm$ 0.079& 2.48 $\pm$ 0.076 & 2.60 $\pm$ 0.13 & 0.62 $\pm$ 0.014 \cr
40000 & 5.78 $\pm$ 0.035& 2.63 $\pm$ 0.34 & 2.84 $\pm$ 0.14 & 0.63 $\pm$ 0.014 \cr
50000 & 6.86 $\pm$ 0.026& 2.63 $\pm$ 0.022 & 3.07 $\pm$ 0.15 & 0.63 $\pm$ 0.014 \cr

\cr
\hline
\hline
\end{tabular}
\end{center}
\end{table*}

\subsection{Grain Rotational Disruption}
According to the theory described in Section \ref{sec:RATs}, to calculate the grain disruption size by RATD, we first need to find the strength and mean wavelength of the radiation field in the protostellar envelope.

%In WC87, $r_{1}$ is the radial distance where the dust temperature $T=T_{\rm sub}=1500\K$, which is $\sim 2\times 10^{15}cm\approx 157 AU$ and $n_{1}$ is the density at $r_{1}$ \citep{1986ApJ...310..207W}. Thus, $r_{1}$ is the inner boundary of the dust cocoon produced by sublimation.

The column density of gas obscuring the massive protostar at a distance $r$ from the inner radius of the dust cocoon, $r_{\rm in}$, is given by 
\bea
N_{\rm H}(r)&=&\int_{r_{\rm in}}^{r} n_{\rm H}(r')dr'\\
&=&\frac{n_{\rm in}r_{\rm in}}{p-1}\left[1- \left(\frac{r}{r_{\rm in}}\right)^{-p+1}\right],\nonumber\\
&\simeq &1.5\times 10^{23}n_{\rm in,8}r_{\rm in,2}\nonumber\\
&&\times\left(\frac{1}{p-1}\right)\left[1-\left(\frac{r}{r_{\rm in}}\right)^{-p+1}\right]\cm^{-2},\nonumber
\ena
where $p=3/2$ (see Eq. \ref{eq:nH}), $r_{\rm in}=r_{\rm sub}$ (Eq. \ref{eq:rsub}), $n_{\rm in,8}=n_{\rm in}/10^{8}\cm^{-3}$ and $r_{\rm in,2}=r_{\rm in}/100\AU$. 

The wavelength-dependent dust extinction is described by $A_{\lambda,\star}=1.086\tau(\lambda)$, and the visual extinction of the central protostar is related to the column density as $A_{V,\star}/N_{\rm H}=R_{V}/(5.8\times 10^{21}\cm^{-2})$ (see \citealt{2011piim.book.....D}). 

The visual extinction measured from the protostar to a radial distance $r$ in the envelope is then given by
\bea \label{eq:Avs-r}
A_{V,\star}(r)&=&\left(\frac{N_{\rm H}(r)}{5.8\times 10^{21}\cm^{-2}}\right)R_{V},\nonumber\\
&=&\frac{A_{V,in}}{p-1}\left[1-\left(\frac{r}{r_{\rm in}}\right)^{-p+1}\right],~~~\label{eq:AV_r}
\ena
where $A_{V,in}=n_{\rm in}r_{\rm in}R_{V}/(5.8\times 10^{21}\cm^{-2})\approx 129n_{\rm in,8}r_{\rm in,2}R_{V}/5$, and $A_{V,\star}(r)=0$ for $r<r_{\rm in}$.

%The visual extinction measured from the cloud surface inward, $A_{V}$, can be calculated using the visual extinction from the protostar, $A_{V,\star}$, as follows
%\bea \label{eq:Avs-Av}
%A_{V}(r)&=&A_{V,\star}(r\gg r_{\rm in})-A_{V,\star}(r)\nonumber\\
%&=&\frac{A_{V,1}}{p-1}\left(\frac{r}{r_{\rm in}}\right)^{-p+1},
%\ena
%and $A_{V}(r<r_{\rm in})=A_{V,1}/(p-1)$.

Due to dust extinction, the intensity of the stellar radiation field decreases but its mean wavelength increases with visual extinction $A_{V,\star}$. Following \cite{Hoang.2021}, the radiation strength of the reddened radiation field at $A_{V,\star}$ from the source can be described by
\bea
U=\frac{U_{\rm source}}{1+c_{1}A_{V,\star}^{c_{2}}}=\frac{U_{\rm in}}{1+c_{1}A_{V,\star}^{c_{2}}}\left(\frac{r}{r_{\rm in}}\right)^{-2},\label{eq:U_AV_star}
\ena
where $U_{\rm source}$ is the strength of the radiation field (e.g., direct stellar radiation or dust emission from the hot shell) at radial distance $r$ in the absence of dust extinction, $U_{\rm in}=L_{\star}/(4\pi r_{\rm in}^{2}cu_{\rm ISRF})$ is the radiation strength at $r=r_{\rm in}$, and $c_{1},c_{2}$ are the fitting parameters. The mean wavelength of the attenuated stellar spectrum is
\bea
\bar{\lambda}=\bar{\lambda}_{\rm source}(1+c_{3}A_{V,\star}^{c_{4}}),\label{eq:wavemean_AV_star}
\ena
where $c_{3}$ and $c_{4}$ are the fitting parameters.

As in \cite{Hoang.2021}, we perform a least chi-square fitting of $U/U_{\rm source}$ using Equation (\ref{eq:U_AV_star}) and $\bar{\lambda}/\bar{\lambda}_{\rm source}$ using Equation (\ref{eq:wavemean_AV_star}) to their numerical values to obtain the best-fit parameters. The best-fit parameters and their uncertainties are shown in Table \ref{tab:coefficient}. 

\begin{figure}
\includegraphics[scale=0.5]{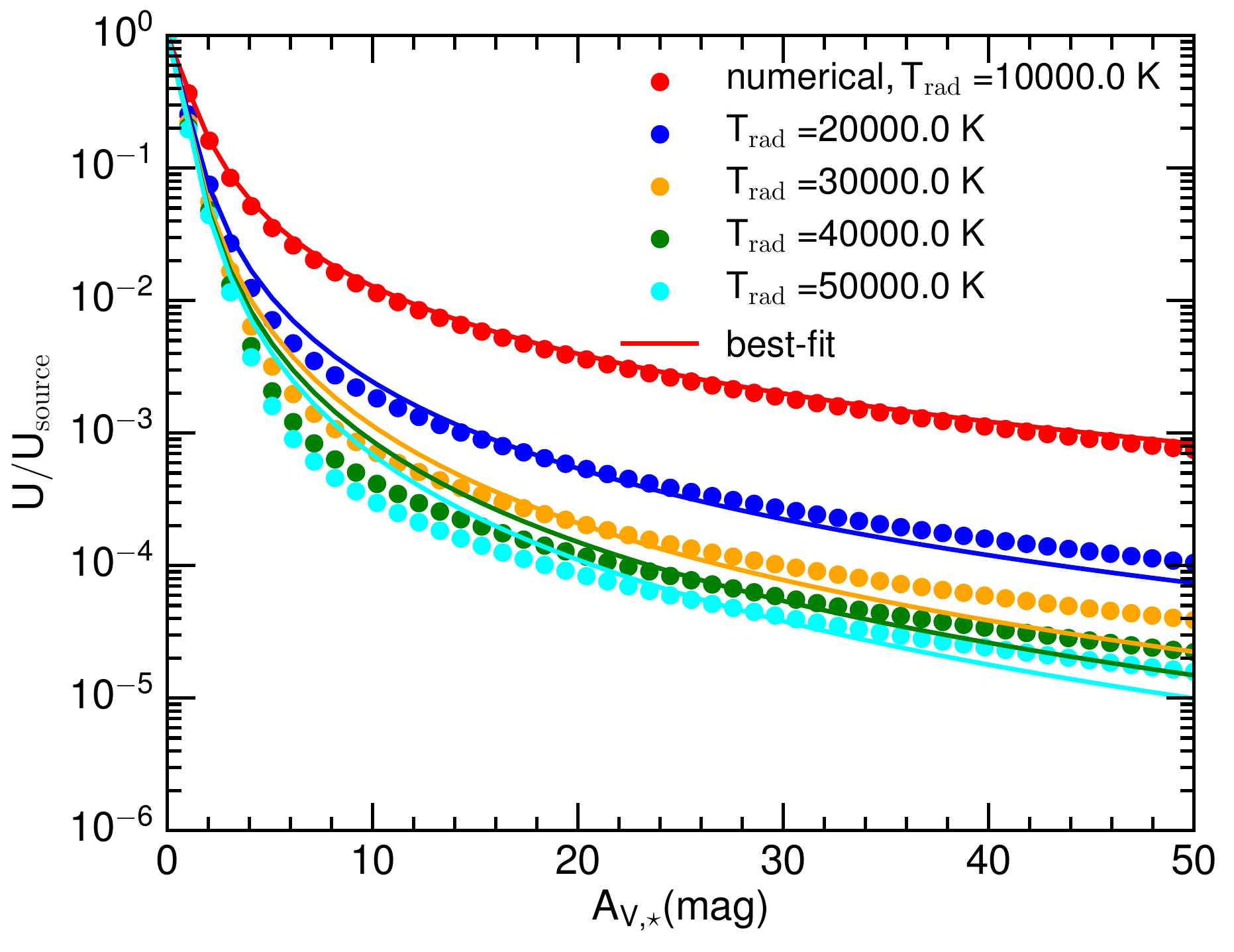}
\includegraphics[scale=0.5]{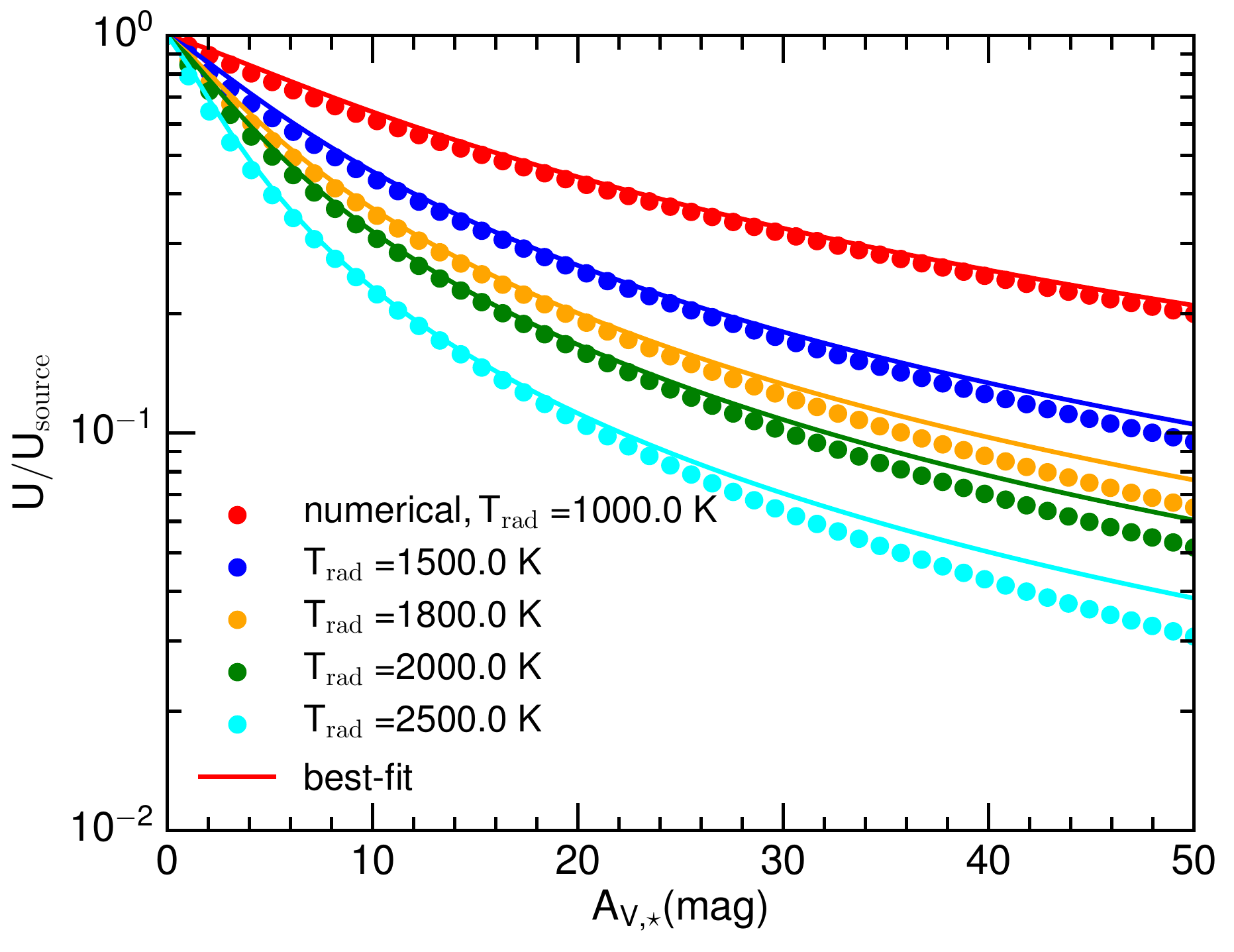}
\caption{Decrease of the radiation strength due to dust extinction for the stellar radiation and the emission from hot dust. Most of radiation from the hot stars ($>90\%$) are absorbed within $A_{V}\sim 1$, producing a hot dust thin shell. Radiation from the hot dust shell can penetrate deeper into the envelope (lower panel).}
\label{fig:U_AV}
\end{figure}

Figure \ref{fig:U_AV} shows the decrease of $U$ with $A_{V,\star}$ for both direct stellar radiation (upper panel) and IR radiation from the hot dust shell (lower panel). As shown, more than $99\%$ of radiation energy from hot stars are absorbed within a thin layer of $A_{V}<5$. This is understandable because such hot stars emit mostly at UV wavelengths and $A_{\rm UV}> 2A_{V}$. Thus, the radiation strength of hot protostars decreases as $U(A_{V})/U_{\rm source}< e^{-2A_{V}}< 10^{-2}$ at $A_{V}>2$. For the radiation field of the hot dust shell, the decrease of $U$ is much slower than in the case of hot stars due to the visual extinction at longer wavelengths.

Assuming the gas-dust thermal equilibrium and using Equation (\ref{eq:U_AV_star}), one obtains the gas temperature as a power-law
\bea
T_{\gas}=T_{\rm in}\left(\frac{r}{r_{\rm in}}\right)^{-q}(1+c_{1}A_{V,\star}^{c2})^{-q/2},\label{eq:Tgas_r}
\ena
where $T_{\rm in}= T_{d,0}U_{\rm in}^{1/(4+\beta)}\K$ is the grain temperature at $r_{\rm in}$ and $q=2/(4+\beta)$ where $T_{d,0}$ is the grain temperature at $U=1$, with $T_{d,0}\approx 16.4\K$ and $\beta\approx 2$ for silicates. The assumption of gas-dust thermal equilibrium for the protostellar core is invalid in the photodissociation region (PDR) around high-mass protostars where gas heating by photoelectric effect is important. 

Following \cite{Hoang.2021}, the disruption size at visual extinction $A_{V,\star}$ from the source is,
\bea
a_{\rm disr}
&=&\left(\frac{3.2n_{\rm H}\sqrt{2\pi m_{\rm H}kT_{\rm gas}}}{3\gamma u_{\rm rad}\bar{\lambda}^{-2}}\right)^{1/2}\left(\frac{S_{\rm max}}{\rho}\right)^{1/4}(1+F_{\rm IR})^{1/2}\nonumber\\
&\simeq& 0.35 \hat{\rho}^{-1/4}S_{\max,7}^{1/4}\left(\frac{U_{\rm in,6}}{n_{\rm in,8}T_{1,2}^{1/2}}\right)^{-1/2}\left(\frac{\bar{\lambda}_{\rm source}}{1.2\mum}\right)\nonumber\\
&&\times(1+c_{1}A_{V,\star}^{c_{2}})^{(1-q/4)/2}(1+c_{3}A_{V,\star}^{c_{4}})\nonumber\\
&&\times\left(\frac{r}{r_{\rm in}}\right)^{(2-p-q/2)/2}(1+F_{\rm IR})^{1/2}\mum,
\label{eq:adisr_AV_star}
\ena
where $U_{\rm in,6}=U_{\rm in}/10^{6}$. 

The RATD effect occurs only if $a_{\rm disr}<a_{\rm trans,\star}=\bar{\lambda}/2.5$ because grains larger than $a_{\rm trans,\star}$ have $\omega_{\rm RAT}$ decreasing with $a$ (Eq. \ref{eq:omega_RAT2}). The maximum size for RATD is given by Equation (\ref{eq:adisr_up}),
\bea
a_{\rm disr,max}
&=&\frac{3\gamma u_{\rm rad}\bar{\lambda}}{64n_{\rm H}\sqrt{2\pi m_{\rm H}kT_{\rm gas}}}\left(\frac{S_{\rm max}}{\rho}\right)^{-1/2}(1+F_{\rm IR})^{-1}\nonumber\\
&\simeq&0.7\hat{\rho}^{1/2}S_{\max,7}^{-1/2}\left(\frac{U_{\rm in,6}}{n_{\rm in,8}T_{1,2}^{1/2}}\right)\left(\frac{\bar{\lambda}_{\rm source}}{1.2\mum}\right)
\nonumber\\
&&\times(1+c_{1}A_{V,\star}^{c_{2}})^{-(1-q/4)}(1+c_{3}A_{V,\star}^{c_{4}})\nonumber\\
&&\times\left(\frac{r}{r_{\rm in}}\right)^{-(2-p-q/2)}(1+F_{\rm IR})^{-1}~\mum.\label{eq:adisr_up_star}
\ena 

Due to a high gas density of the protostellar core, Equation (\ref{eq:FIR}) implies $F_{\rm IR}\ll 1$. Thus, the $1+F_{\rm IR}$ term in the above equations can be ignored, so that we can obtain analytical results for the disruption size $a_{disr}$.

The above equations can be used to obtain the disruption size by the direct stellar radiation and the infrared emission from the hot dust shell (see Eq. \ref{eq:u_td}). To check the validity of our analytical results, we will numerically calculate the alignment size (disruption size) using $\omega_{\rm RAT}$ from Equation (\ref{eq:omega_RAT0}) where $\Gamma_{\rm RAT}$ is numerically computed using Equation (\ref{eq:GammaRAT_num}) and apply the criteria for grain alignment (disruption) for the stellar radiation using the reddening law (Eq. \ref{eq:U_star_red}). The obtained results are referred to as numerical results. Note that for numerical calculations, the IR damping is considered (cf. analytical results).

We use Equations (\ref{eq:adisr_AV_star} and \ref{eq:adisr_up_star}) to calculate the disruption by both the stellar radiation and by the IR emission from the hot dust shell. For the stellar radiation, we take $T_{\rm source}=T_{\star}$ and $T_{\rm source}=T_{\rm shell}$ for the IR emission from the hot shell. 

\section{Numerical Results}\label{sec:results}
\subsection{Grain disruption size}
We first calculate $a_{\rm disr}$ as a function of the radius $r$ from the central massive star for the different model parameters shown in Table \ref{tab:model}. We fix the protostar mass of $M_{\star}=60M_{\odot}$ and $100M_{\odot}$ and the total stellar luminosity of $L_{\rm tot}$. The accretion rate is varied between $10^{-4}-5\times 10^{-3}M_{\odot}\yr^{-1}$, which results in the different density profiles (Eq. \ref{eq:nH}). 

We assume that large grains have a composite structure and adopt $r_{0}=0.1\mum$, yielding the typical tensile strength of $S_{\rm max}=10^{5}\erg\cm^{-3}$. This assumption is consistent with the popular paradigm of grain evolution in which grains grow in dense clouds due to coagulation, resulting in composite/fluffy grain structure (\citealt{1990ARA&A..28...37M}). We also explore the possibilities that large grains are made of smaller monomers ($r_{0}<0.1\mum$) that have a larger tensile strength of $S_{\rm max}=10^{7}\erg\cm^{-3}$. We consider only silicate grains, but our present theory can be generalized for other dust compositions because RATs are insensitive to dust compositions (\citealt{2007MNRAS.378..910L}; \citealt{2019ApJ...878...96H}).

We calculate the disruption size for both the stellar radiation and infrared emission from the hot dust shell. We find that disruption by stellar radiation is negligible because the UV radiation is significantly decreased at $A_{V,\star}\sim 5$. In the following, we only discuss the results induced by the hot dust shell.

\begin{table}
%\begin{center}
\caption{Model Parameters}\label{tab:model}
\begin{tabular}{l l} \hline\hline
{Parameters} & Values\cr
\hline\\
Stellar mass, $M_{\star}$& $60M_{\odot}, 100M_{\odot}$ \cr
Stellar luminosity, $L_{\rm tot}$ & $5\times 10^{5}L_{\odot},10^{6}L_{\odot}$\cr
Stellar temperature, $T_{\star}$& $30,000\K$ \cr
Accretion rate, $\dot{M}$& $10^{-4}-5\times 10^{-3} M_{\odot}\rm yr^{-1}$ \cr
Inner radius, $r_{\rm in}$ & $r_{\rm sub}$\cr
Outer radius, $r_{2}$ & \cr
%Shell mass, $M_{\rm shell}$ & $200-300M_{\odot}$\cr 
Grain tensile strength & $S_{\max}$\cr
\hline
\hline
\end{tabular}
%\end{center}
\end{table}

%\begin{figure}
%\includegraphics[width=0.5\textwidth]{nH_Tgas_M100_Lstar1e6.pdf}
%\caption{Density and temperature profile of the envelope around a massive protostar.}
%\end{figure}

\begin{figure*}
\includegraphics[width=0.5\textwidth]{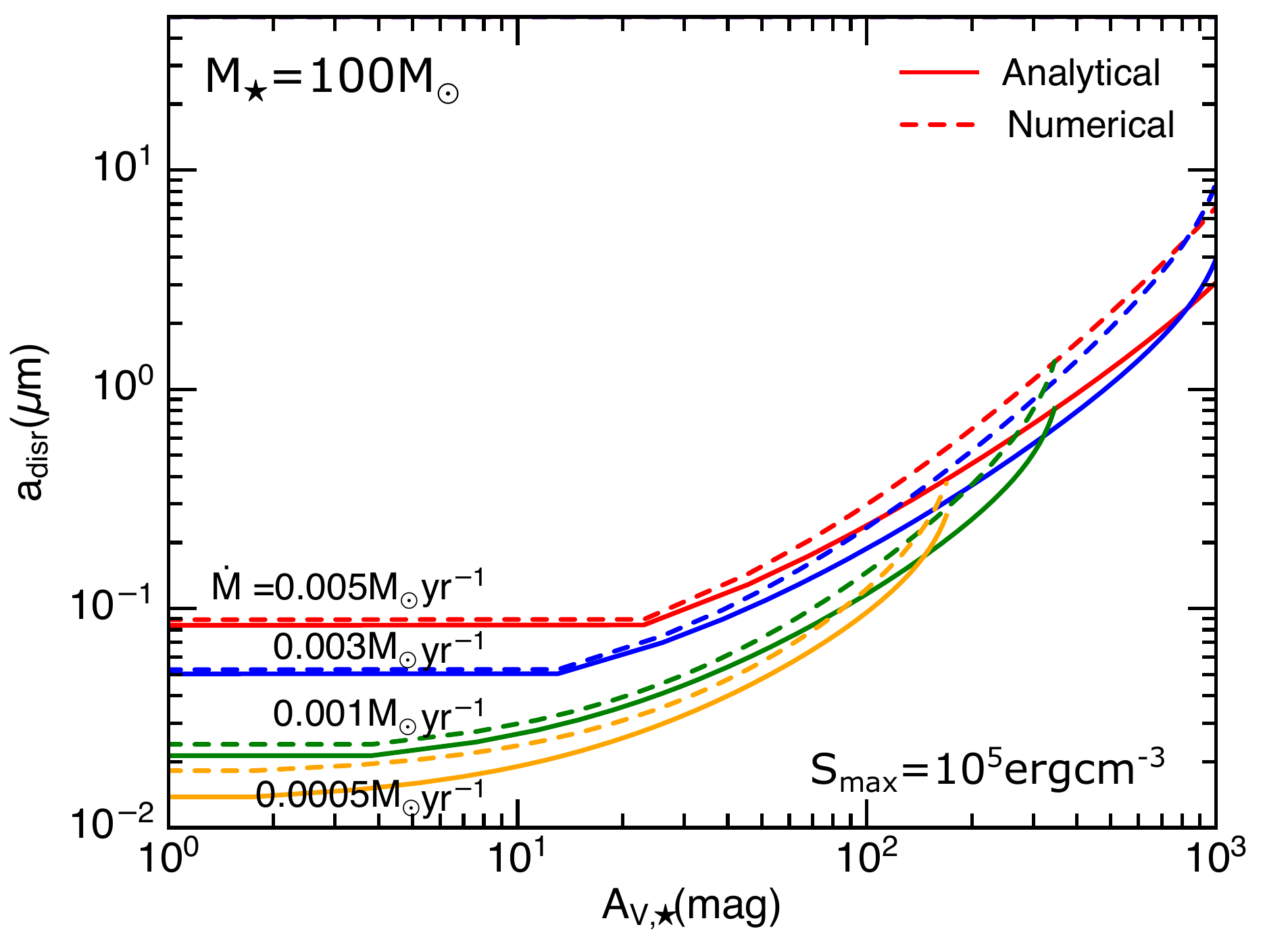}
\includegraphics[width=0.5\textwidth]{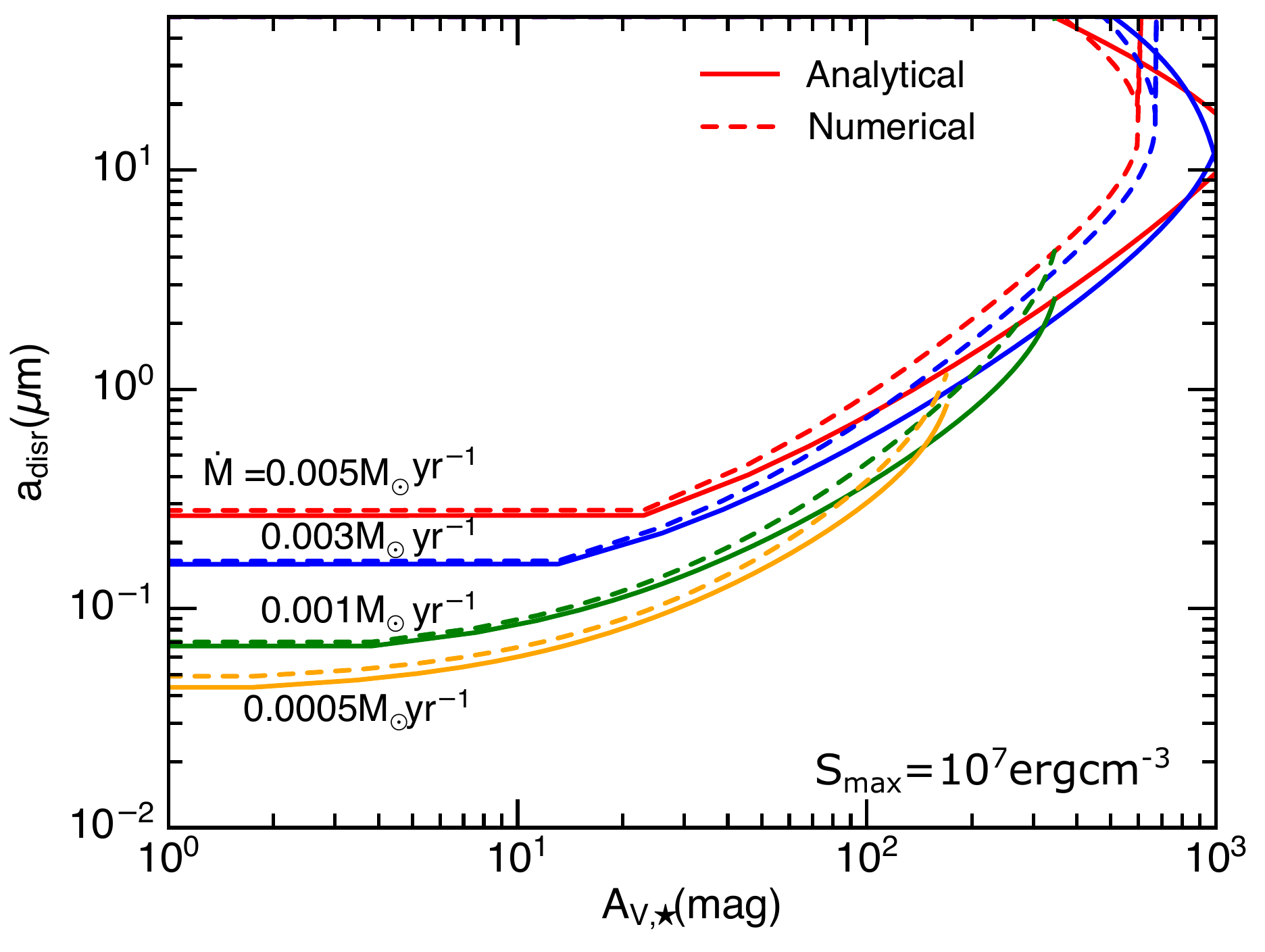}
\includegraphics[width=0.5\textwidth]{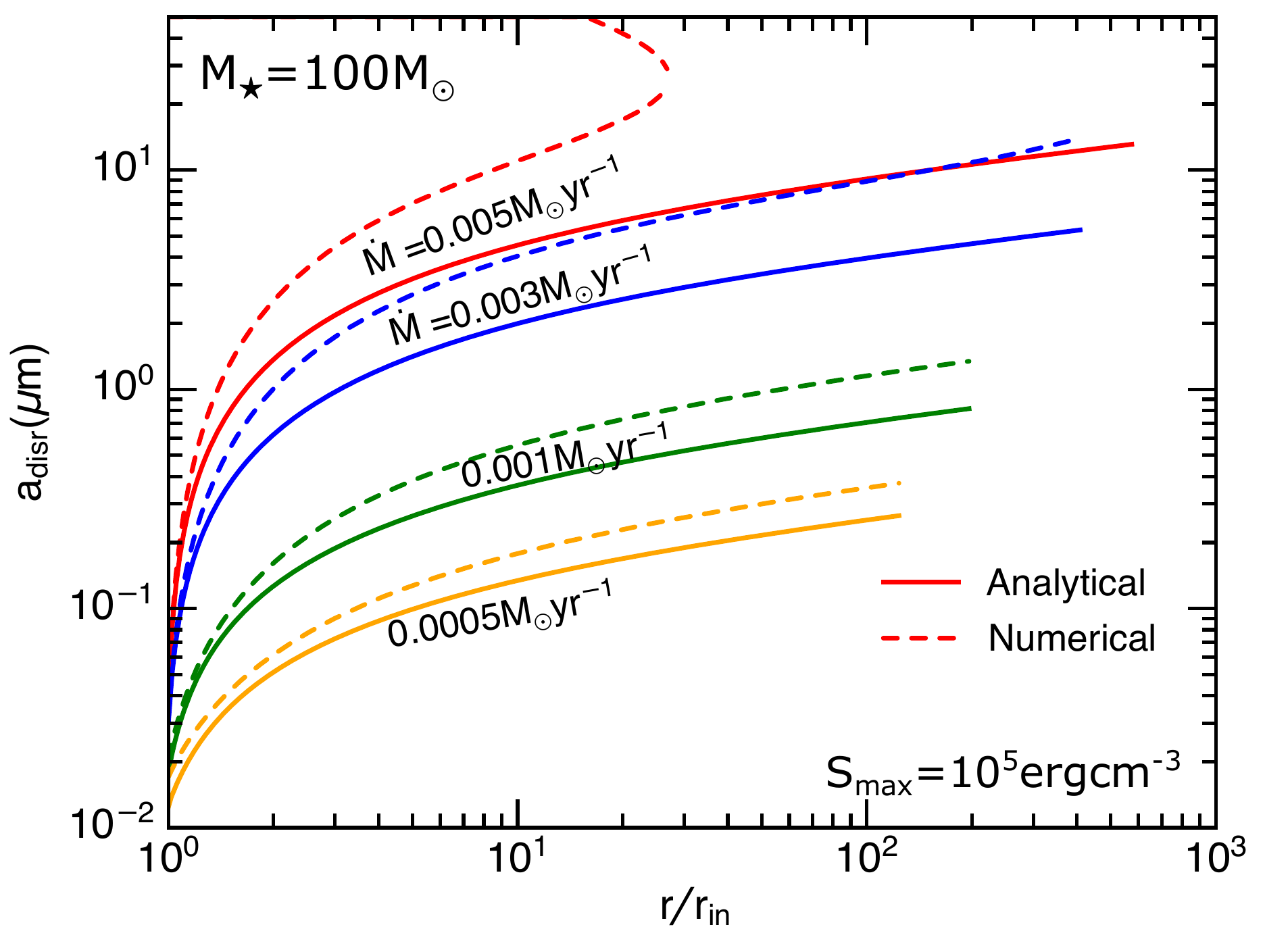}
\includegraphics[width=0.5\textwidth]{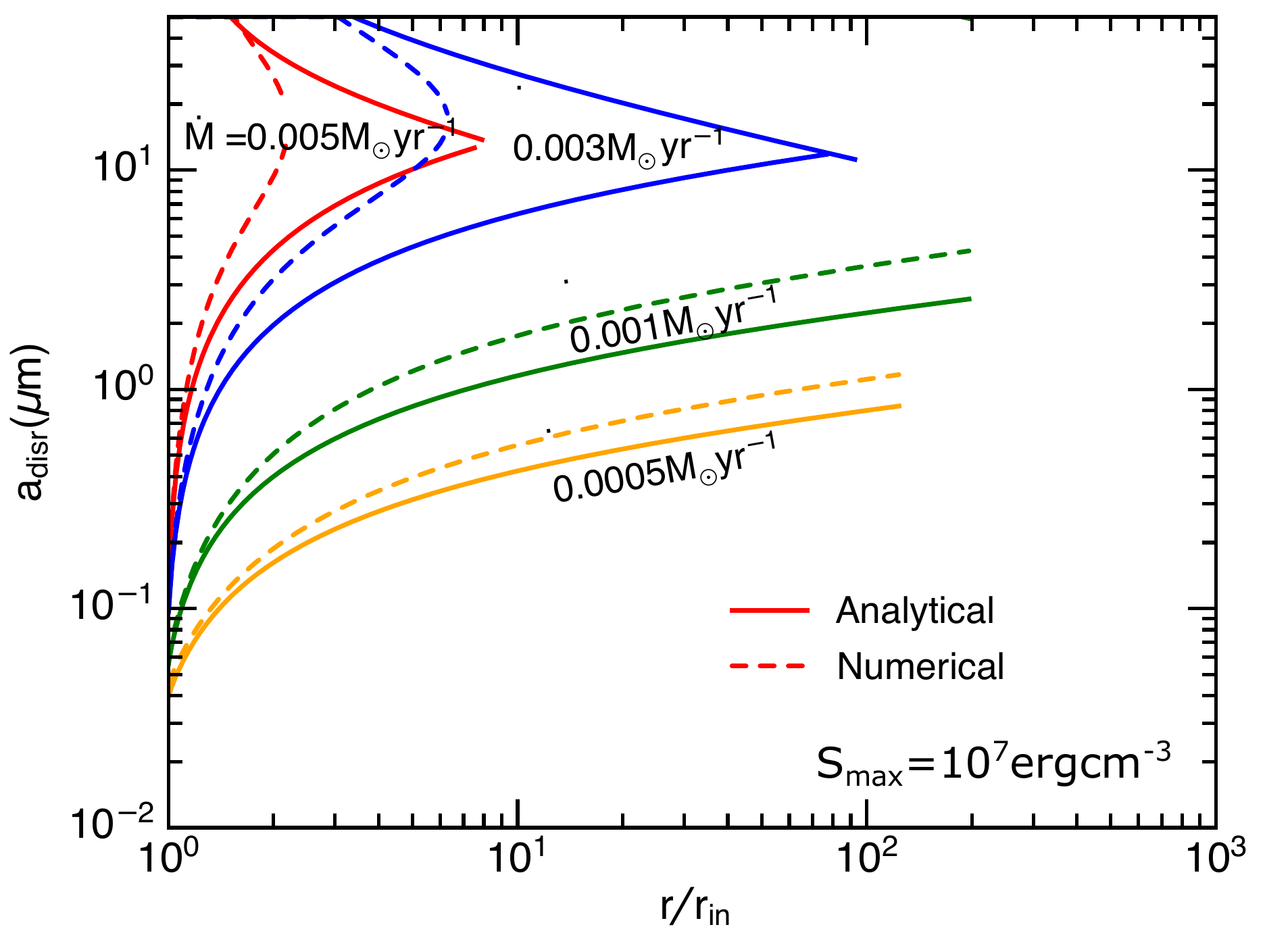}
\caption{Variation of the disruption size obtained from analytical equations and numerical methods, as a function of visual extinction for different accretion rates $\dot{M}=5\times 10^{-3}-10^{-4}M_{\odot}\yr^{-1}$, assuming $S_{\max}=10^{5}\erg\cm^{-3}$ (left panel) and $10^{7}\erg\cm^{-3}$ (right panel). In each panel, the lower and upper lines show $a_{\rm disr}$ and $a_{\rm disr,max}$, respectively, and the space constrained by $a_{\rm disr},a_{\rm disr,max}$ determine the disruption zone. RATD is considered effective if $a_{\rm disr}<a_{\rm max, orig}$. RATD efficiency increases toward the central star due to increase of the radiation field. RATD efficiency is stronger for smaller accretion rate.}
\label{fig:adisr_AV_hotdust_high}
\end{figure*}

Figure \ref{fig:adisr_AV_hotdust_high} shows the variation of $a_{\rm disr}$ and $a_{\rm disr,max}$ obtained from our analytical formulae (solid lines) with numerical results (dashed lines) as functions of $A_{V,\star}$, assuming the typical tensile strength of $S_{\max}=10^{5}\erg\cm^{-3}$ for large composite grains of $a>0.1\mum$ (see Eq. \ref{eq:Smax}). One can see that the grain disruption size is larger for higher accretion rate $\dot{M}$ due to higher gas density (Eq. \ref{eq:nH}) which results in stronger rotational damping. The disruption size decreases rapidly toward the central protostar due to increasing radiation intensity.

For very high accretion rate of $\dot{M}>10^{-3}M_{\odot}\yr^{-1}$, grain disruption is inefficient (i.e., $a_{\rm disr}>1\mum$) in the outer region due to large gas density. For accretion rates of $\dot{M}\lesssim10^{-3}M_{\odot}\yr^{-1}$, micron-sized grains ($a> 1\mum$) can be disrupted even in the outer envelope, therefore, dust in the entire envelope is completely processed by RATD and their size distribution returns to that of the diffuse ISM with $a_{\rm max}\sim 0.1\mum$ (see Figure \ref{fig:adisr_AV_hotdust_high}).

%\begin{figure}
%\includegraphics[width=0.5\textwidth]{adisr_AV_Smax1_massivestar.pdf}
%\caption{Same as figure \ref{fig:adisr_AV_hotdust_high}, but for the disruption by the direct stellar radiation. Disruption only occurs within a small region of $A_{V}\sim 10$.}
%\label{fig:adisr_AV_star_high}
%\end{figure}

%Figure \ref{fig:adisr_AV_star_high} shows the disruption size due to the stellar radiation. The disruption only occurs near the star with $A_{V,\star}<30$, much smaller than the active region induced by infrared emission from the hot dust shell.

\subsection{Radiation Pressure Opacity in the presence of RATD}
Due to RATD, the grain size distribution is modified from the original size distribution which is plausibly that of the prestellar cores. Because grain growth is efficient in dense cores such as in dense protostellar cores and protostellar disks of very high densities of $n_{\H}\sim 10^{7}-10^{8}\cm^{-3}$ given by Equation (\ref{eq:nH}), we assume the maximum size of the original dust is $a_{\rm max, orig}=10\mum$ (\citealt{2013MNRAS.434L..70H}. In the presence of RATD, the maximum size $a_{\rm max}$ is determined by $\min(a_{\rm disr},a_{\max,orig}$ because $a_{\rm disr,max}>0.5\mum$. 

\begin{figure*}
\includegraphics[width=0.5\textwidth]{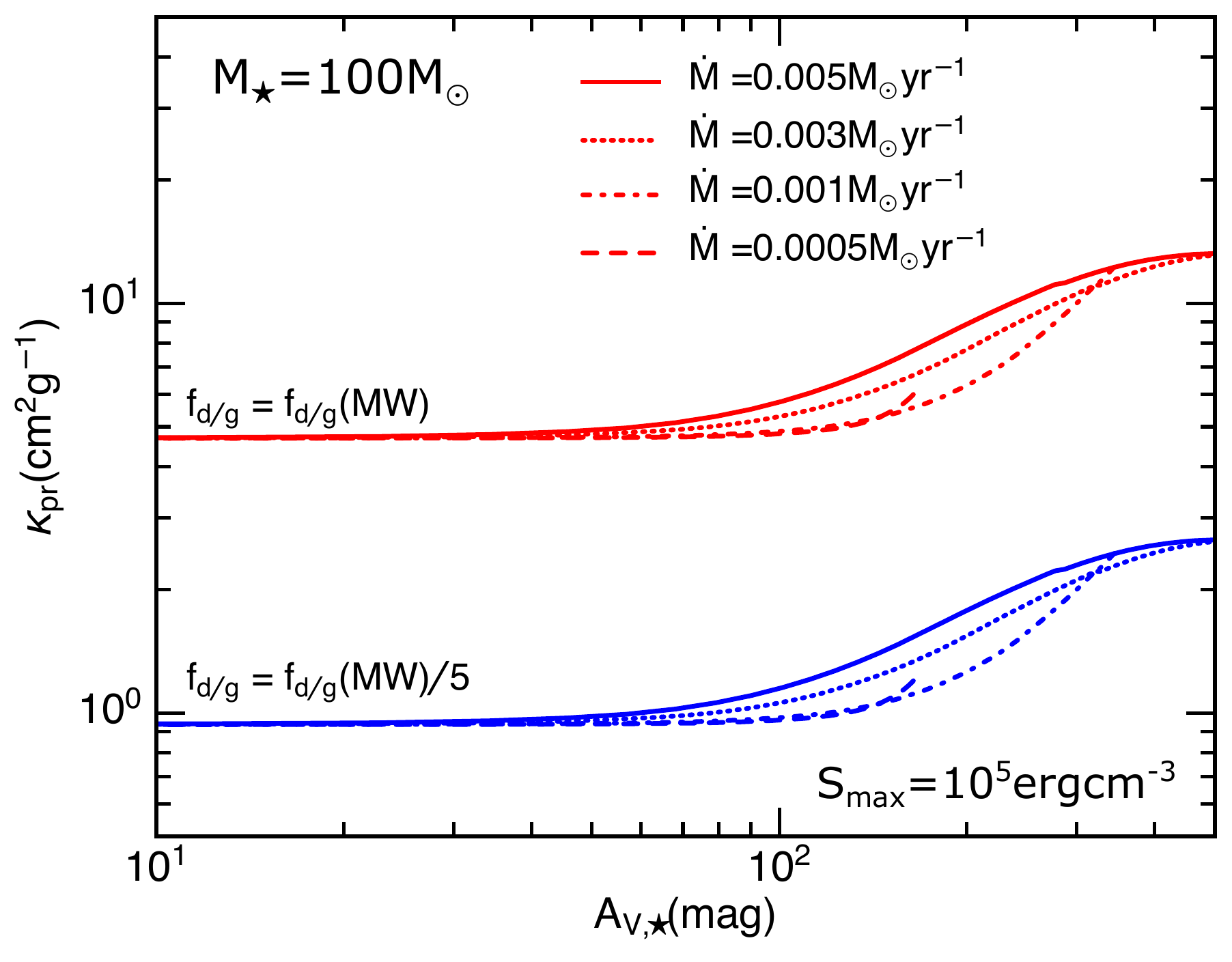}
\includegraphics[width=0.5\textwidth]{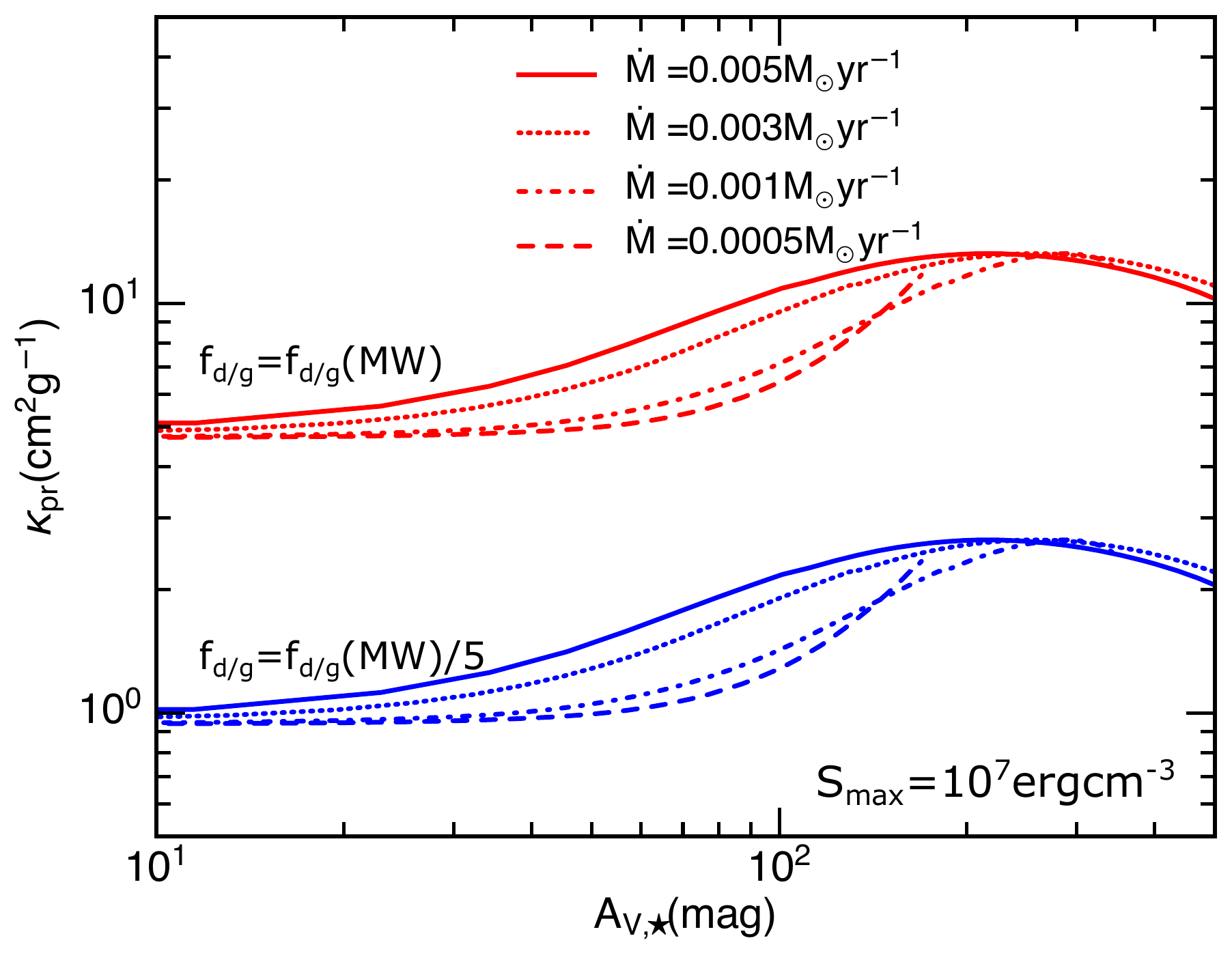}
\caption{Radiation pressure opacity per dust mass as a function of $A_{V}$ from the central star for $M_{\star}=100M_{\odot}$ and different accretion rates, for $f_{d/g}=f_{d/g}(\rm MW)=0.01$ (red lines) and $f_{d/g}=f_{d/g}(\rm MW)/5$ (blue lines), assuming $S_{\rm max}=10^{5}$ and $10^{7}\erg\cm^{-3}$. The opacity $\kappa_{\rm pr}$ decreases rapidly from the outer envelope ($A_{V,\star}\sim 300-500$) toward the central star and becomes smaller than its maximum value at the outer envelope by a factor $\sim 3$.}
\label{fig:kappa_pr_AV}
\end{figure*}

Using $a_{\rm disr}(r)$ from RATD, we compute the radiation pressure opacity using Equation (\ref{eq:kappad}) with $u_{\lambda}$ given by the hot dust shell.

Figure \ref{fig:kappa_pr_AV} shows the decrease of the radiation pressure opacity with the visual extinction for two values of the tensile strength, $S_{\max}=10^{5}\erg\cm^{-3}$ (upper panel) and $S_{\max}=10^{7}\erg\cm^{-3}$ (right panel) for two cases of the standard (red lines) and reduced (blue lines) dust-to-gas ratio, $f_{d/g}$. For larger accretion rate of $\dot{M}=5\times 10^{-3}M_{\odot}\yr^{-1}$, disruption does not occur at large $A_{V}$ and the maximum opacity of $\kappa_{\rm pr}\sim 13.2 \cm^{2}\g^{-1}$ at the outer envelope ($A_{V,\star}\sim 500$). It decreases to the minimum value of $\kappa_{\rm pr}\sim 4.7\cm^{2}\g^{-1}$ at $A_{V}\sim 50$. For lower accretion rate, $\dot{M}\lesssim 10^{-3}M_{\odot}\yr^{-1}$, the disruption occurs even at larger $A_{V}$ and $\kappa_{\rm pr}$ decreases significantly.
\begin{figure*}
\includegraphics[width=0.5\textwidth]{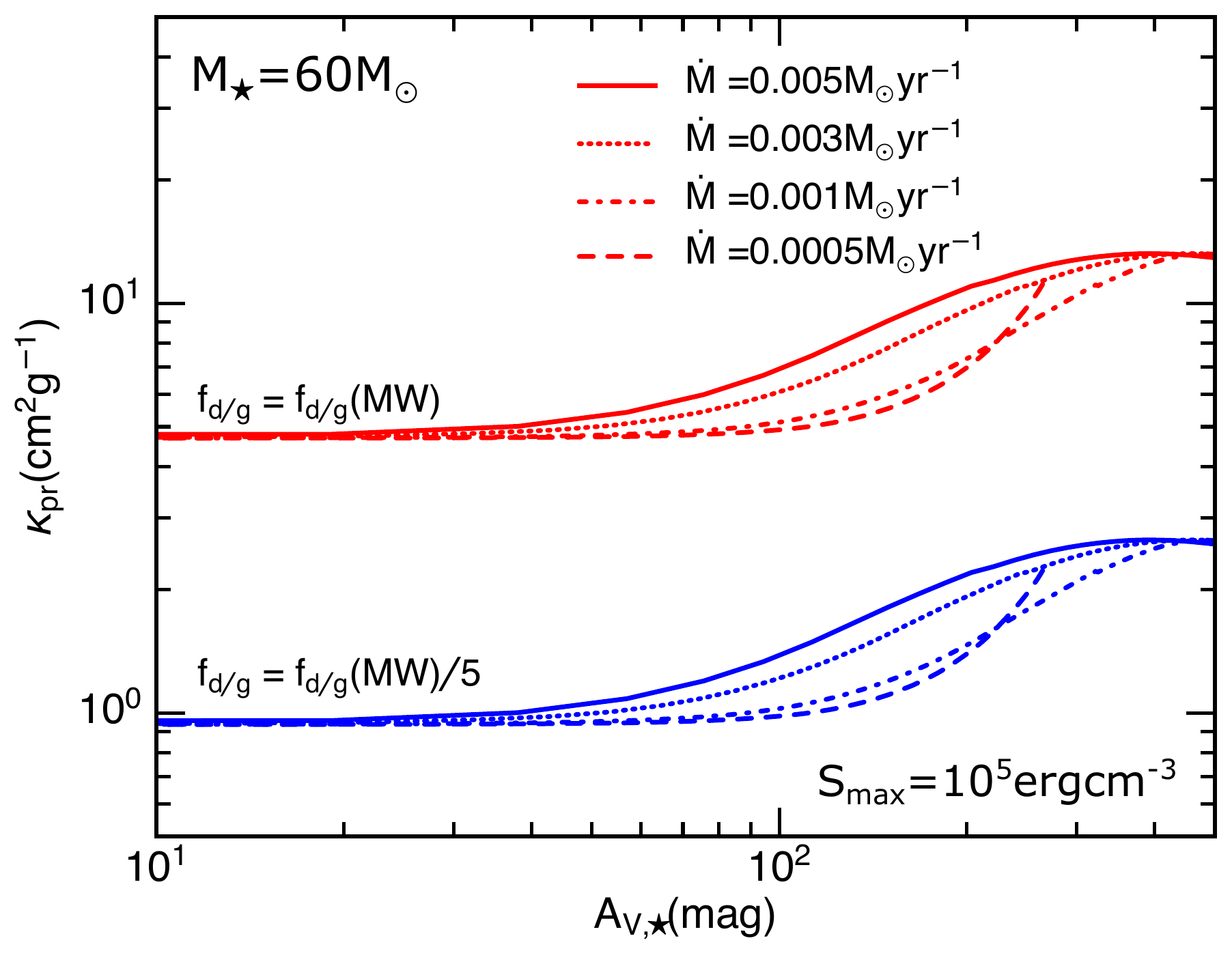}
\includegraphics[width=0.5\textwidth]{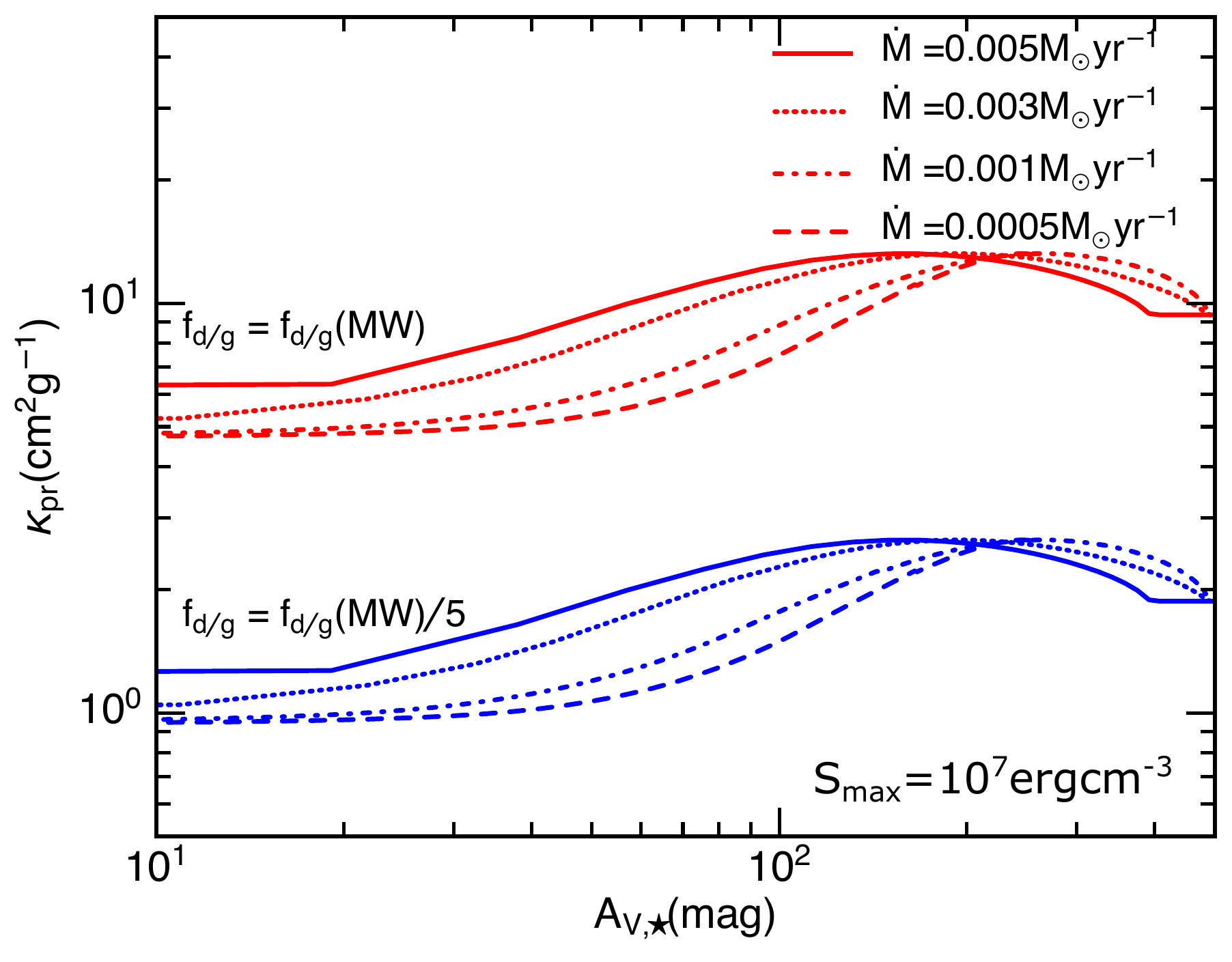}
\caption{Same as Figure \ref{fig:kappa_pr_AV} but for $M_{\star}=60M_{\odot}$ and $L_{\rm tot}=5\times 10^{5}L_{\odot}$.}
\label{fig:kappa_pr_AV_M60}
\end{figure*}

\subsection{Implications for Massive Star Formation}
Infalling gas in the envelope is subject to gravitational force and radiation pressure on dust. The ratio of radiation pressure force on dust to gravity is given by
\bea
\Lambda = \frac{F_{\rm rad}}{F_{\rm grav}}&=&\frac{\kappa_{\rm pr}L_{\rm tot}/(4\pi r^{2}c)}{GM_{\star}/r^{2}}=\frac{L_{\rm tot}\kappa_{\rm pr}}{4\pi G M_{\star}c}\nonumber\\
&=&3.9\frac{f_{d/g}}{0.01}\left(\frac{\kappa_{\rm pr}}{5\g\cm^{-2}}\right)\left(\frac{L_{\rm tot}}{10^{6}L_{\odot}}\right)\left(\frac{M_{\star}}{100M_{\odot}}\right)^{-1},\label{eq:Gamma}
\ena
where $\kappa_{\rm pr}$ is given by Equation (\ref{eq:kappa_pr}). 

To form a massive star via accretion, the gravity must exceed the radiation pressure, i.e., $\Lambda<1$, which corresponds to
\bea
\frac{L_{\rm tot}}{M_{\star}}<\frac{L_{\rm Edd,d}}{M_{\star}}=2564.1\frac{f_{d/g}}{0.01}\left(\frac{5\g\cm^{-2}}{\kappa_{\rm pr}}\right) \frac{L_{\odot}}{M_{\odot}},\label{eq:LM_rp}
\ena
where $L_{\rm Edd,d}$ is the Eddington luminosity, maximum luminosity that the envelope is not blown away. However, massive stars of $M_{\star}>20M_{\odot}$ already have $L_{\rm tot}/M_{\star}>2500$. Therefore, if the dust opacity is similar to the standard ISM, accretion cannot form stars above $20M_{\odot}$. 

%The mass of the shell taken from the sublimation front is
%\bea
%M_{\rm shell}(r)=\int_{r_{\rm in}}^{r}\rho_{\gas}(r) 4\pi r^{2} dr=\frac{8\pi \rho_{\rm in}}{3}r_{\rm in}^{3/2}(r^{3/2}-r_{\rm in}^{3/2}),\label{eq:Mshell}
%\ena
%where $\rho_{\rm in}$ is the mass density at $r_{\rm in}$.

Using $\kappa_{\rm pr}(r)$ obtained from the previous section, as shown in Figure \ref{fig:kappa_pr_AV}, we now compute $\Lambda=F_{\rm rad}/F_{\rm grav}$ at several distances $r$ due to RATD, assuming the different values of $\dot{M}$, $M_{\star}=100M_{\odot}$, same parameters as in WC87 (Figure 3).

\begin{figure*}
\includegraphics[width=0.5\textwidth]{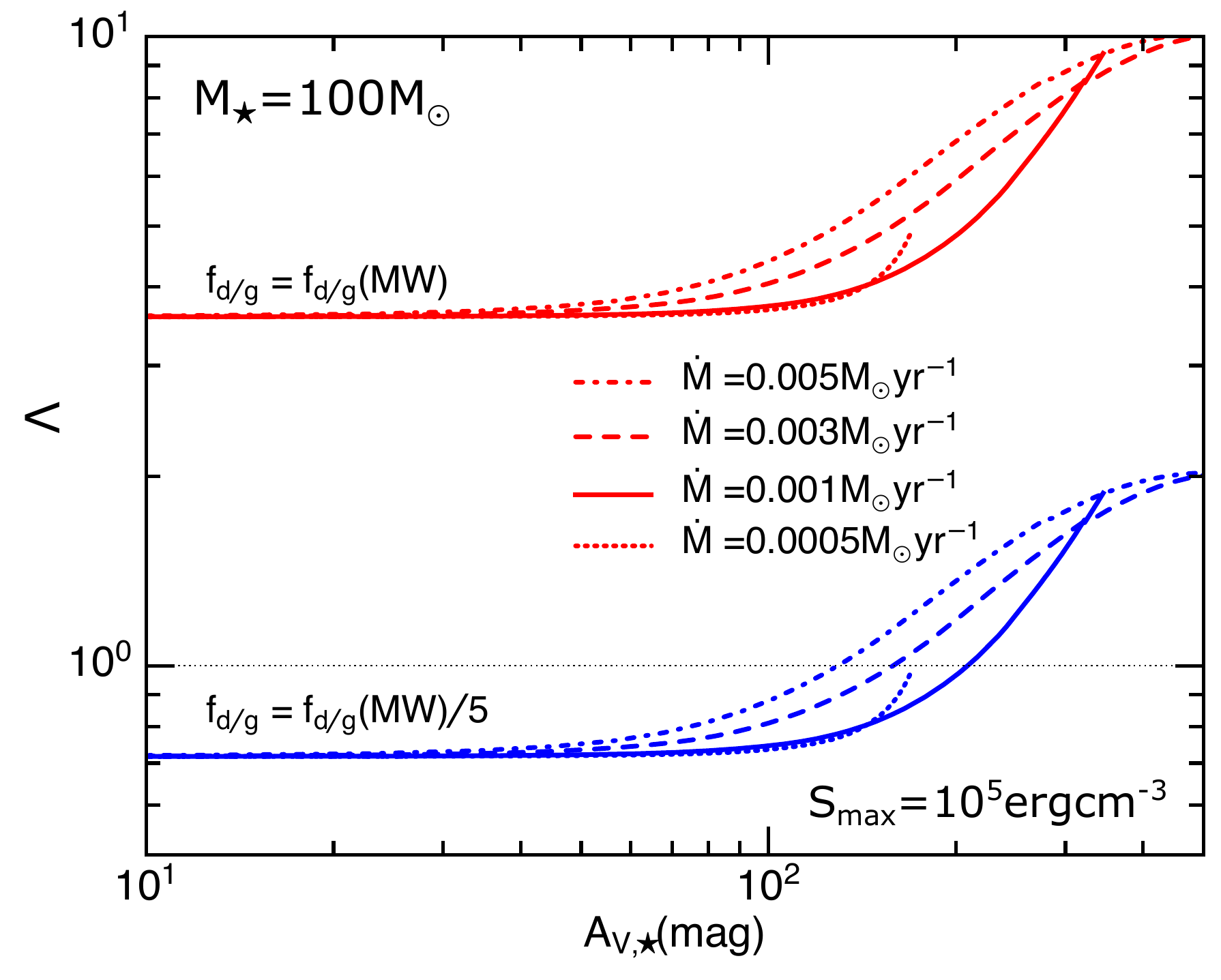}
\includegraphics[width=0.5\textwidth]{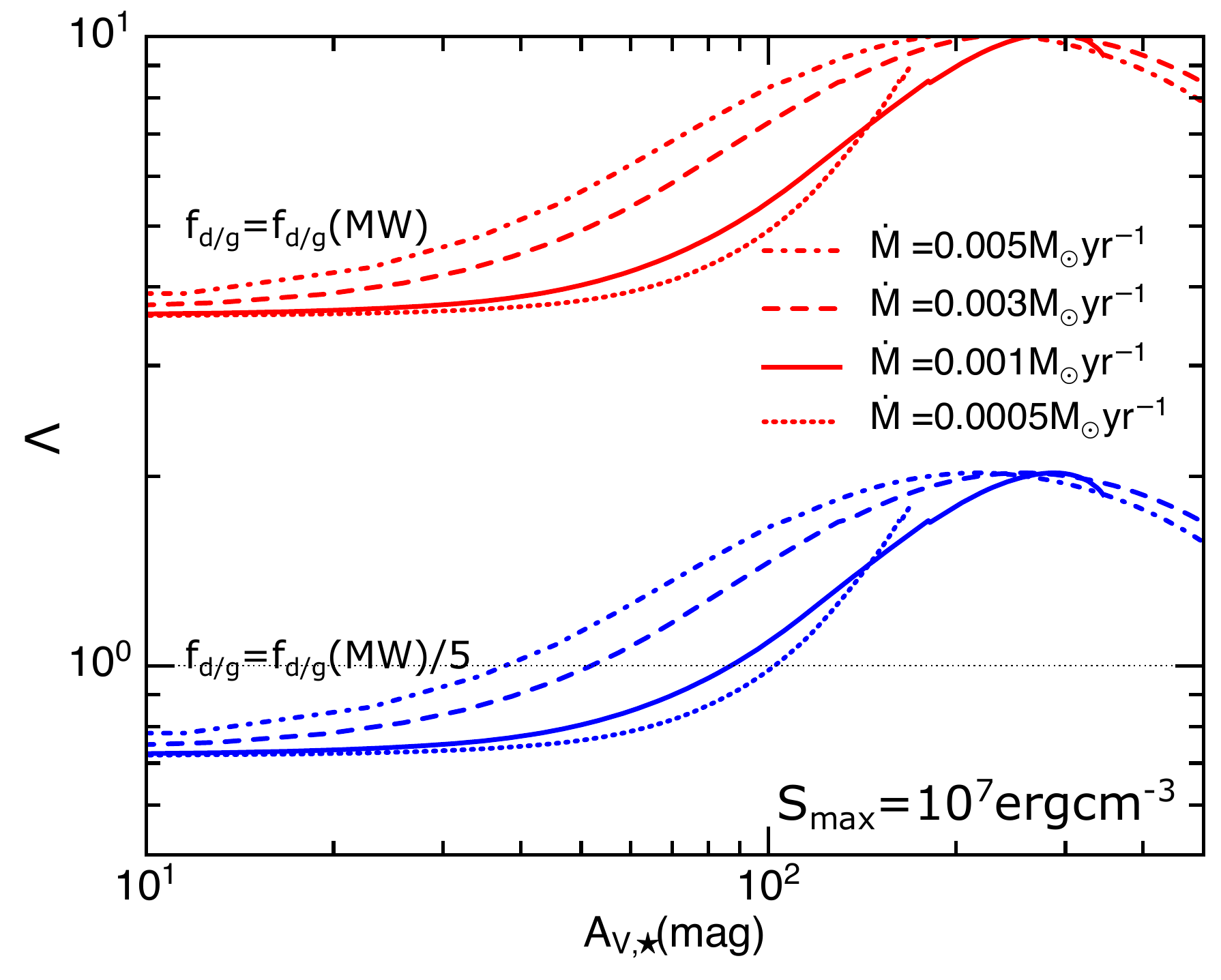}
\caption{The ratio of radiation pressure vs. gravity, $\Lambda$, for the standard $f_{g/d}$ (upper panel) and $f_{d/g}=f_{d/g}(\rm MW)/5$ assuming $S_{\max}=10^{5}\erg\cm^{-3}$.}
\label{fig:Gamma_pr}
\end{figure*}

\begin{figure*}
\includegraphics[width=0.5\textwidth]{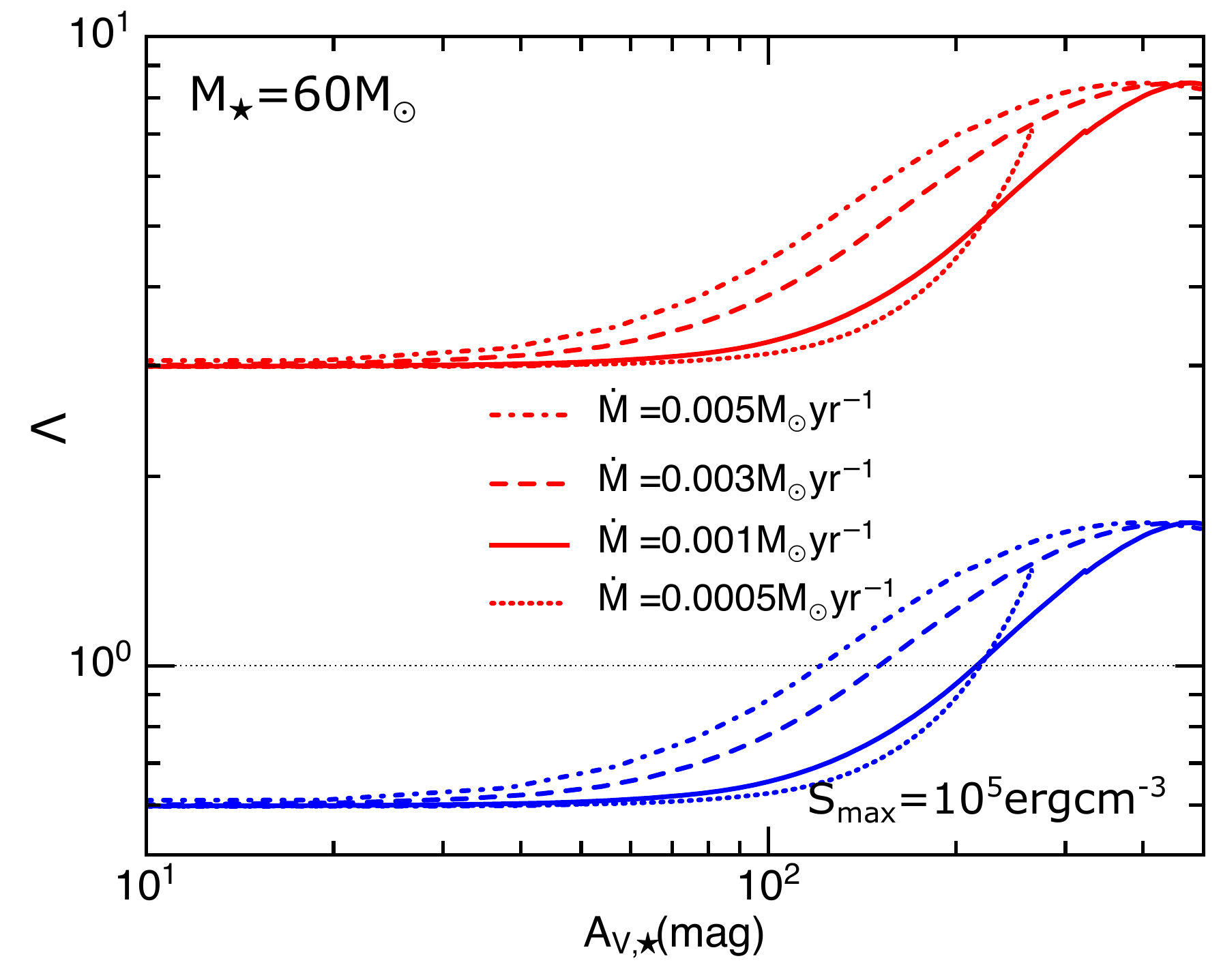}
\includegraphics[width=0.5\textwidth]{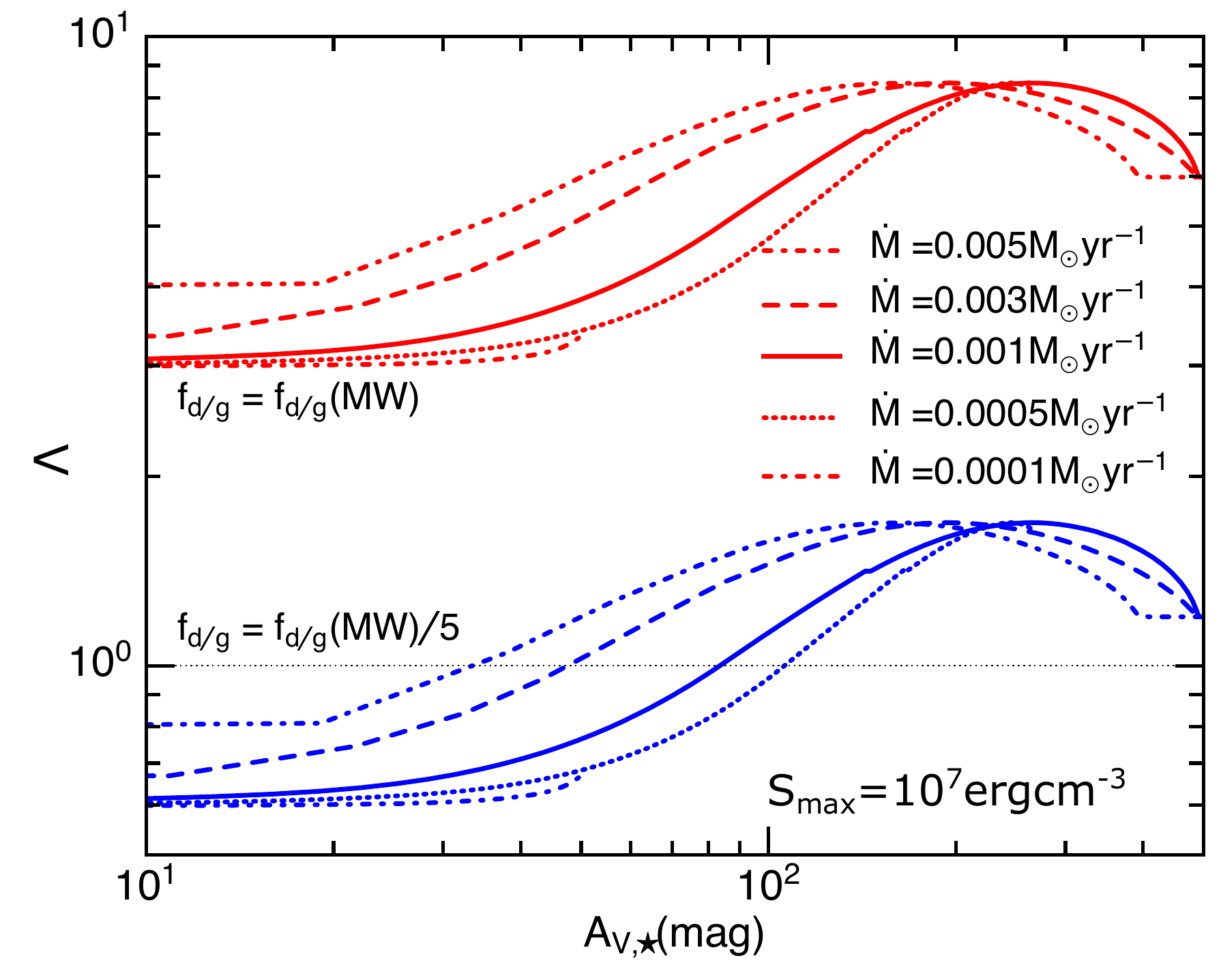}
\caption{Same as Figure \ref{fig:Gamma_pr} but for for $M_{\star}=60M_{\odot}$ and $L_{\rm tot}=5\times 10^{5}L_{\odot}$.}
\label{fig:Gamma_pr_M60}
\end{figure*}

Figure \ref{fig:Gamma_pr} shows $\Lambda$ as a function of the radial distance for the different tensile strengths , assuming the typical $M_{\star}=100M_{\odot}$ for $f_{d/g}=f_{d/g}(\rm MW)=0.01$ and $f_{d/g}=f_{d/g}(\rm MW)/5$. One can see that $\Lambda$ decreases rapidly with $r$ due to the effect of RATD. A smaller accretion rate $\dot{M}$ can reduce $\Lambda$ at a large distance, and a higher $\dot{M}$ only reduces $\Lambda$ in the inner region near the sublimation front. For the case of $f_{d/g}=f_{d/g}(\rm MW)/5$, one can see that $\Lambda<1$ for $A_{V,\star}<100-200$. For a low accretion rate of $\dot{M}\lesssim 0.001M_{\odot}\yr^{-1}$, one has $\Lambda<1$ for $S_{\max}=10^{5}\erg\cm^{-3}$. Therefore, the radiation pressure can be overcome if grains in dense clouds are composite. For more compact grains with $S_{\max}=10^{7}\erg\cm^{-3}$, grain disruption is less efficient, and one has $\Lambda<1$ only for $A_{V,\star}<100$. Moreover, the decrease of $f_{d/g}$ will enable RATD at large distances from the source, which reduces the dust opacity.

Figure \ref{fig:Gamma_pr_M60} shows the same results but for lower stellar mass and luminosity with $M_{\star}=60M_{\odot}$ and $L_{\rm tot}=5\times 10^{5}L_{\odot}$. The main features are similar to those in Figure \ref{fig:Gamma_pr}, but $\Lambda$ is lower due to smaller luminosity $L_{\rm tot}$.

\section{Discussion}\label{sec:disc}

\subsection{Dust properties in massive protostellar envelopes}
Dust properties are crucially important for understanding the role of radiation pressure feedback in massive star formation. Unfortunately, the dust properties are not well constrained in protostellar envelopes. Large grains are expected to be present in the prestellar cores due to grain growth implied by theory
\citealt{2013MNRAS.434L..70H}) and observations (\citealt{2010Sci...329.1622P}; \citealt{Lefevre:2020fw}; \citealt{2013ApJ...763...55R}; \citealt{Ysard:2013fn}). However, under intense radiation field from massive protostars, dust properties are expected to change. Recent advances in dust physics reveal that dust grains could be disrupted by the RATD mechanism (\citealt{Hoang:2019da}; \citealt{2020Galax...8...52H}) beyond the sublimation front where the temperature is much below $T_{\rm sub}$.

Using the RATD mechanism, we showed that micron-sized grains are rapidly disrupted into smaller ones by radiative torques. Therefore, the abundance of large grains is reduced in the inner envelope of protostars, while the abundance of small grains increases toward the central protostar. The effect is efficient for realistic accretion rates $\dot{M}$, and only inefficient for very high accretion rate of $\dot{M}>5\times 10^{-3}M_{\odot}\yr^{-1}$.

%We note that RATD is inefficient in the protostellar disk interior where the radiation is shielded and gas density is extreme of $n_{\H}>10^{10}\cm^{-3}$ \citep{Tung:2020ew}.

Modeling of observed spectral energy density from the dense regions around O stars usually infer the grain size distribution similar to the MRN with $a_{\max}=0.25\mum$ (\citealt{1986ApJ...310..207W}; \citealt{1987ApJ...319..850W}; \citealt{1990ApJ...354..247C}). This is unexpected because grains grow efficiently to micron sizes in dense cores where massive stars form. Therefore, RATD is a plausible mechanism to resolve this tension.

%Interestingly, AME is detected strongly in a compact region of $\sim 100$ pc which is thought to be a massive-star forming region \cite{2018ApJ...862...20M} where thermal dust is significantly reduced \cite{Murphy:2020te}. For the AME region of NGC 4975, the young massive cluster has luminosity equal to $167$ O7 stars, which is $L\sim 10^{9}L_{\odot}$. For this intense radiation, grains can be disrupted out to $100$ pc, which is the size of a GMC \cite{Hoang:2019da}, which implies the reduction of thermal dust and increase of spinning dust. To estimate the effect of RATD on spinning dust emissivity, we find that a typical grain of $a=0.1\mum$ can be disrupted into $N_{nano}=(0.1/0.001)^{3}=10^{6}$ nanoparticles of radius $1$ nm. Since the spinning dust emissivity is proportional to the density of nanoparticles $n_{nano}$, the disruption (even partial disruption) can easily increase the emissivity by a factor of $10^{3}$ as observed by \cite{Murphy:2020te}. 

\subsection{Reduced radiation pressure by RATD and massive star formation}

The radiation pressure on dust is thought to be the major barrier for the accretion by gravity, which prevents the formation of very massive stars (\citealt{1971A&A....13..190L}). Reduction of radiation pressure is required to form massive stars. Previous studies assuming the MRN size distribution (\citealt{1986ApJ...310..207W}; \citealt{1987ApJ...319..850W} suggested that if large grains could be removed and the dust mass is reduced by a factor of 4, then the radiation pressure can be circumvented and form stars of $M_{\star}>20M_{\odot}$. 

We first found that with the realistic grain size distribution in the protostellar envelopes, the radiation pressure opacity is several times larger than obtained with the MRN distribution (see Figure \ref{fig:kappa_pr}), which strengthens the radiation pressure problem. Thanks to RATD, large grains are disrupted into small ones, and dust becomes smaller toward the star. As a result, the radiation pressure opacity in the envelope irradiated by the hot dust shell's radiation is found to decrease toward the central protostar. Accordingly, the ratio of the radiation pressure to the gravitational force decreases toward the central star due to RATD. Nevertheless, to form very massive stars, the dust-to-gas ratio still needs to be reduced by a factor of $\sim 5$. The physical mechanism underlying such dust destruction is unknown. 

\subsection{Effect of hot dust emission on grain alignment}
In addition to rotational disruption (RATD effect), RATs are known to induce grain alignment (\citealt{1997ApJ...480..633D}; \citealt{2007MNRAS.378..910L}; \citealt{Hoang:2008gb}; \citealt{2016ApJ...831..159H}). The relation between RAT alignment and RATD is discussed in detail by \cite{LH:2021}. Observations of RAT alignment are presented in a review by \cite{2015ARA&A..53..501A}. 

Grain alignment by protostellar radiation in dense clouds is studied in detail by \cite{Hoang.2021} for low-mass and high-mass protostars. The authors found that stellar radiation can align grains at large $A_{V,\star}$ from the central source. However, the authors only consider the stellar temperature of $T_{\star}\lesssim 10^{4}\K$ where most stellar radiation energy is concentrated in optical-NIR. We now discuss the alignment by stellar radiation from very massive protostars with $T_{\star}\gtrsim 2\times 10^{4}\K$ for which stellar radiation is substantially absorbed within a narrow region of $A_{V,\star}\lesssim 5$ (see Figure \ref{fig:U_AV}), and the reemission by the hot dust shell plays a more important role in aligning grains in the envelope.

Following \cite{Hoang.2021}, the minimum size of aligned grains (hereafter alignment size) at visual extinction $A_{V,\star}$ from the protostar,
\bea
    a_{\rm align} &=&\left(\frac{4n_{\rm H}T_{\rm gas}}{3\gamma u_{\rm rad}\bar{\lambda}^{-2}}\right)^{2/7}\left(\frac{15m_{\rm H}k^{2}}{4\rho}\right)^{1/7}(1+F_{\rm IR})^{2/7}\nonumber\\
        &\simeq &0.031\hat{\rho}^{-1/7} \left(\frac{U_{\rm in,6}}{n_{\rm in,8}T_{\rm in,2}}\right)^{-2/7}\left(\frac{\bar{\lambda}_{\rm source}}{1.2\mum}\right)^{4/7}  \nonumber\\
    &&\times (1+c_{1}A_{V,\star}^{c_{2}})^{(2-q)/7}  (1+c_{3}A_{V,\star}^{c_{4}})^{4/7}\nonumber\\
&&\times \left(\frac{r}{r_{\rm in}}\right)^{2(2-p-q)/7} (1+F_{\rm IR})^{2/7}\mum,
\label{eq:aalign_AV_star}
\ena
where $\gamma=1$ is adopted for the unidirectional field from the protostar.

%Figure \ref{fig:aalign_AV_star_high} (upper panel) shows the alignment size as a function of $A_{V,\star}$ due to direct stellar radiation from the hot star of $T_{\star}=3\times 10^{3}\K$. The alignment size is rather large for the outer envelope and decreases only toward the central star.
 
\begin{figure}
\includegraphics[width=0.5\textwidth]{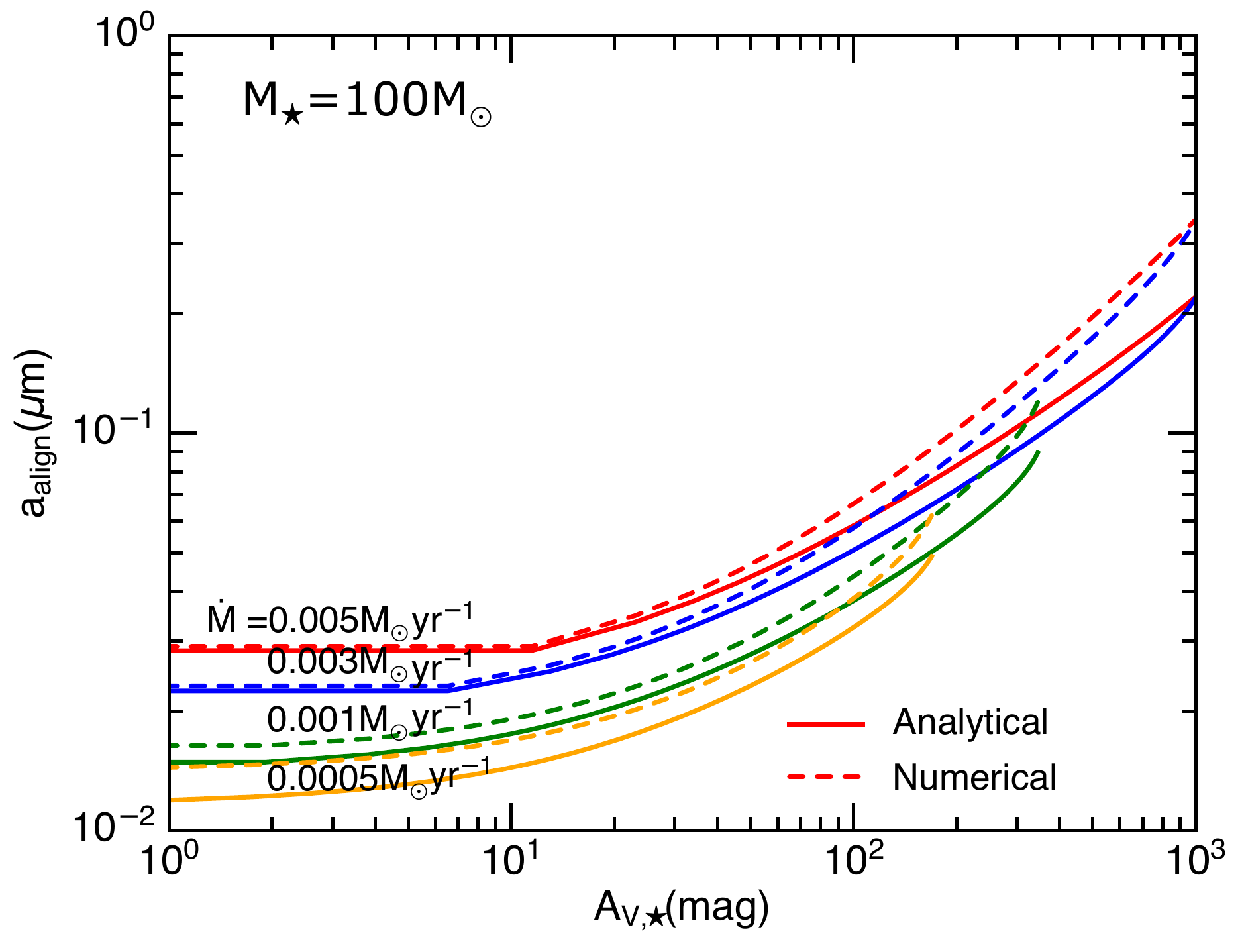}
\caption{Grain alignment size as a function of the visual extinction to the star by reemission from the hot dust, assuming $M_{\star}=100M_{\odot}$ and $L_{\star}=10^{6}L_{\odot}$. Both analytical results (Eq. \ref{eq:aalign_AV_star} with $F_{\rm IR}=0$) and numerical results are shown. The alignment size is slightly larger for numerical results due to the more important contribution of IR damping when the accretion rate is lower.}
\label{fig:aalign_AV_star_high}
\end{figure}

Figure \ref{fig:aalign_AV_star_high} shows the alignment size due to infrared emission from the hot dust shell of $T_{d}=1000\K$ obtained using Equation (\ref{eq:aalign_AV_star}) by setting $F_{\rm IR}=0$ (i.e., analytical results) and numerical results obtained using numerical method (see \citealt{Hoang.2021}). Grains can still be aligned at large distances to the reprocessed stellar radiation and become more efficient when moving toward the star. We also calculate the alignment by direct stellar radiation and find that it is effective only within $A_{V,\star}<10$, i.e., in a much narrow region compared to alignment by hot dust shell. Therefore, near- to mid-IR emission from hot dust could be important for alignment of grains at large $A_{V,\star}$ from hot OB stars with temperatures of $T_{\star}\gtrsim 2 \times 10^{4}\K$.

It is worthy to note that stellar winds might reduce dust mass due to sputtering in the shocked region. However, during the massive star formation, the ram pressure by infalling gas can be much larger than that by the stellar winds because the accretion rate is several orders of magnitude larger than the mass loss rate by stellar winds. Thus, stellar winds are diminished rapidly by infalling gas.

\subsection{Effect of RATD in protostellar disks}
{ To get insight into the effect of RATD in the protostellar envelope on the radiation pressure feedback, we have assumed a spherical collapse toward the central protostar for our numerical calculations. Here we discuss the effect of RATD in the protostellar disks around the massive protostars.

Both observations (see \citealt{Cesaroni.2006}) and numerical simulations (e.g., \citealt{2013ApJ...767L..11K}) establish that massive star formation proceeds with the formation of an accretion disk. The formation of accretion disks is an effective way to overcome radiation pressure barrier because radiation can easily escape along the rotation axis due to its low density. However, the exact radius of accretion disks is uncertain. While hydrodynamics simulations by \cite{Klassen.2016} reveal the rapid formation of large, rotationally-supported disks due to gravity, simulations including stellar feedback and magnetic fields by \cite{Rosen.2020} show that the large scale disks cannot form as expected from the magnetic braking effect (\citealt{Allen.2003}). Misaligned magnetic field and rotation axis as well as non-ideal MHD effects are expected to form a small accretion disk (see \citealt{Wurster.2018} and \citealt{Zhao:2020fl} for recent reviews).

We performed calculations for an extended envelope from $r_{in}=r_{sub}\sim 100-150$ au for $L\sim 10^{6}L_{\odot}$ to $r_{out}\sim 10^{4}$ au (see Eq. \ref{eq:rsub}). For the typical disk radius of $r_{disk}<1000$ au, our results shown in Figure \ref{fig:aalign_AV_star_high} perhaps remain valid for $r>10r_{in}$, but become inapplicable for the inner region of $r<10r_{in}$ because the gas density is much larger. For example, for a protostellar disk formed from hydrodynamic simulations in \cite{Klassen.2016}, the gas density is $n_{\H}\sim 10^{11} \cm^{-3}$ at $\sim 100$ au and $\sim 10^{8}\cm^{-3}$ at $\sim 1000$ au from the central star (see their Figure 11). For a disk from simulations with the turbulence and stellar feedback in \cite{Rosen.2020}, the density is $n_{\H}\sim 10^{8}\cm^{-3}$ at $1000$ au from the star in the absence of the magnetic field. In the presence of magnetic fields, the gas density is much smaller at the same radius (see \citealt{Rosen.2020}). These gas densities are larger than the density given by our density profile (Eq. \ref{eq:nH}), which implies $n_{\H}\sim 10^{10}\cm^{-3}$ for $\dot{M}=0.005 M_{\odot}~yr^{-1}$ at $r\sim 100 $ au. Therefore, our results for the high accretion rate of $\dot{M}=0.005M_{\odot}yr^{-1}$ may be not too far from the results obtained for a realistic protostellar disk.

It is worth to mention that \cite{Tung:2020ew} performed a detailed numerical modeling of RATD for the circumstellar disk of radius $\sim 300$ au around low mass stars of luminosity $L_{\star}\sim 35L_{\odot}$ with the radiative transfer treated by RADMC-3D. The authors found that the RATD is efficient in the disk surface and intermediate layer only, but is inefficient in the disk mid-plane. Therefore, with the massive protostar luminosity of much higher luminosity of $L_{\star}\sim 10^{6}L_{\odot}$, the effect of RATD would be more efficient than in the disk around low and intermediate mass stars. We will carry out a detailed modeling of RATD for protostellar disks around massive stars in a follow-up paper.
}

\section{Summary}\label{sec:summary}
We study the dynamical effects of intense radiation feedback from massive protostars on the dust in the protostellar envelope and explore its implications for massive star formation. Our main findings are summarized as follows:

\begin{enumerate}

\item The dust radiation pressure opacity depends crucially on the grain size distribution (described by the maximum size $a_{\max}$) and the radiation field spectrum.

\item 
We study the effect of radiation force and torques on the dust and find that grain rotational disruption by RATs (i.e., RATD) is always faster than acceleration by radiation pressure. Thus, dust properties in intense radiation are radically different from the dust in starless cores prior to the onset of star formation.

\item
We find that large, micron-sized grains of porous structures, which are expected in dense protostellar envelopes due to grain evolution, are rapidly disrupted into smaller ones by IR radiation from the hot dust shell heated by the intense stellar radiation. This effect transforms the original dust size distribution into the type of the diffuse ISM. The disruption is efficient in the dust cocoon and increases toward the inner region.

\item
We calculate dust radiation pressure opacity using the size distribution determined by RATD. The resulting IR radiation pressure opacity decreases with distance to the central star and can be reduced by a factor of $\sim 3$ compared to the original opacity without RATD, whereas the UV opacity increases significantly. Therefore, MIR photons from the hot dust shell can escape from the envelope more efficiently. 

\item Radiation pressure feedback is less efficient compared to the realistic model without dust disruption by RATs. However, it still requires the reduction of the dust mass by a factor of $sim 5$ to form very massive stars. Thus, the radiation pressure barrier is indeed a challenge for massive star formation in the spherical collapse scenario.

\item Dust properties in the massive star-forming core are radically different from the standard ISM dust due to rotational disruption by RATs. An accurate understanding of radiation pressure on massive star formation requires a detailed study of dust physics accounting for the new effects.

\end{enumerate}

\acknowledgments
{ We are grateful to the anonymous referee for a thorough and useful report.} T.H. acknowledges the support by the National Research Foundation of Korea (NRF) grants funded by the Korea government (MSIT) through the Mid-career Research Program (2019R1A2C1087045).
%We are grateful to the anonymous referee for useful comments and suggestions. 

\appendix
\section{Grain Rotational Damping}\label{sec:appendix}
The well-known damping process for a rotating grain is sticking collision with gas species (atoms and molecules), followed by their thermal evaporation. Thus, for a gas with He of $10\%$ abundance, the characteristic damping time is
\bea
\tau_{\gas}&=&\frac{3}{4\sqrt{\pi}}\frac{I}{1.2n_{\rm H}m_{\rm H}
v_{\rm th}a^{4}}=\frac{2\sqrt{\pi}\rho a}{6n_{\H}\sqrt{2kT_{\gas}m_{\H}}}\nonumber\\
&\simeq& 2.6a_{-5}\hat{\rho}\left(\frac{10^{6}\cm^{-3}}{n_{\H}}\right)\left(\frac{100\K}{T_{\gas}}\right)^{1/2}~{\rm yr},~~
\ena
where $I=8\pi \rho a^{5}/15$ is the grain inertia moment of spherical grain of effective radius $a$, $v_{\rm th}=\left(2k_{\B}T_{\rm gas}/m_{\rm H}\right)^{1/2}$ is the thermal velocity of a gas atom of mass $m_{\rm H}$ in a plasma with temperature $T_{\gas}$ and density $n_{\H}$ (\citealt{1996ApJ...470..551D}; \citealt{2009ApJ...695.1457H}). The gas damping time is estimated for spherical grains, and we disregard the factor of unity due to grain shape. 

Infrared (IR) photons emitted by the grain carry away part of the grain's angular momentum, resulting in the damping of the grain rotation. For strong radiation fields or not very small sizes, grains can achieve equilibrium temperature, such that the IR damping coefficient (see \citealt{1998ApJ...508..157D}) can be calculated as
\bea
F_{\rm IR}\simeq 0.12\left(\frac{U_{6}^{2/3}}{a_{-5}}\right)
\left(\frac{10^{6} \cm^{-3}}{n_{\H}}\right)\left(\frac{100 \K}{T_{\gas}}\right)^{1/2}.\label{eq:FIR}
\ena 

Other rotational damping processes include plasma drag, ion collisions, and electric dipole emission. These processes are mostly important for polycyclic aromatic hydrocarbons (PAHs) and very small grains of radius $a<0.01\mum$ (\citealt{1998ApJ...508..157D}; \citealt{Hoang:2010jy}; \citealt{2011ApJ...741...87H}). Thus, the total rotational damping rate by gas collisions and IR emission can be written as
\bea
\tau_{\rm damp}^{-1}=\tau_{\gas}^{-1}(1+ F_{\rm IR}).\label{eq:taudamp}
\ena

For strong radiation fields of $U\gg 1$ and not very dense gas, one has $F_{\rm IR}\gg 1$. Therefore, $\tau_{\rm damp}\sim \tau_{\gas}/F_{\IR}\sim a_{-5}^{2}U^{2/3}$, which does not depend on the gas properties. In this case, the only damping process is caused by IR emission.

%--------------adding references-----------------------------------
%\bibliographystyle{/Users/thiemhoang/Dropbox/Papers3/apj}
% or other styles: mcbride,plain, abbrv, acm, alpha, apalike, apj
%\bibliography{/Users/thiemhoang/Dropbox/Papers3/cites_paperApJ}
\bibliography{ms.bbl}

\end{document}